%% file: main.tex
\documentclass[a4paper,11pt]{article}
\usepackage{aaskaiid}
\usepackage{orcidlink}
\usepackage[utf8]{inputenc}
\usepackage[T1]{fontenc}
\usepackage{amsmath}
\usepackage{graphicx}
\usepackage[english]{babel}
\usepackage{url}
\usepackage{hyperref}
\usepackage{caption}
\usepackage{natbib}
\usepackage{array}
\newcommand\Tstrut{\rule{0pt}{3ex}}         
\newcommand\Ttwostrut{\rule{0pt}{5ex}}         
\newcommand\Bstrut{\rule[-2ex]{0pt}{0pt}}   

\title{Opening new parameter space windows on galaxy/AGN co-evolution with SKA radio continuum surveys}
\ShortTitle{A new radio window on galaxy/AGN co-evolution}

\author[1]{Isabella Prandoni\orcidlink{0000-0001-9680-7092}}
\author[2]{Mark T. Sargent\orcidlink{0000-0003-1033-9684}}
\author[1,3]{Matteo Bonato\orcidlink{0000-0001-9139-2342}}
\author[4]{Dharam V. Lal\orcidlink{0000-0001-5470-305X}}
\author[5]{Mahdiyar Mousavi-Sadr\orcidlink{0000-0002-5170-2534}}
\author[6]{Masoumeh Ghasemi-Nodehi\orcidlink{0000-0001-6113-0317}}
\author[7]{Nicholas Seymour\orcidlink{0000-0003-3506-5536}}
\author[5]{Fatemeh S. Tabatabaei\orcidlink{0000-0002-0377-0970}}
\author[8]{Gianfranco De Zotti\orcidlink{0000-0003-2868-2595}}
\author[1,3]{Ivano Baronchelli\orcidlink{0000-0003-0556-2929}}
\author[1,3]{Elisabetta Liuzzo\orcidlink{0000-0003-0995-5201}}
\author[1,3]{Nicola Marchili\orcidlink{0000-0002-5523-7588}}
\author[1,3]{Marcella Massardi\orcidlink{0000-0002-0375-8330}}
\author[1,3]{Rosita Paladino\orcidlink{0000-0001-9143-6026}}

\ShortName{I. Prandoni, M. Sargent, M. Bonato et al.} 

\affiliation[1]{INAF-IRA, Via P. Gobetti 101, 40129 Bologna, Italy}
\emailAdd{isabella.prandoni@inaf.it}
\affiliation[2]{Institute of Physics, Laboratory of Astrophysics, \'Ecole Polytechnique F\'ed\'erale de Lausanne (EPFL), Observatoire de Sauverny, Versoix CH-1290, Switzerland}
\emailAdd{mark.sargent@epfl.ch}
\affiliation[3]{ALMA-ARC, Via P. Gobetti 101, 40129 Bologna, Italy}
\emailAdd{matteo.bonato@inaf.it}
\affiliation[4]{National Centre for Radio Astrophysics - TIFR, Post Box 3, Ganeshkhind P.O., Pune 411007, India}
\affiliation[5]{Institute for Research in Fundamental Sciences (IPM), School of Astronomy, Tehran, Iran}
\affiliation[6]{Xinjiang Astronomical Observatory, CAS, 150 Science-1 Street, Urumqi 830011, China}
\affiliation[7]{International Centre for Radio Astronomy Research, Curtin University, Bentley, WA 6102, Australia}
\affiliation[8]{INAF, Osservatorio Astronomico di Padova, Vicolo Osservatorio 5, I-35122, Padova, Italy}

\abstract{In this chapter we provide an overview of the science enabled by the SKAO, focusing on galaxy/AGN co-evolution studies. In particular we discuss a number of radio continuum `reference' surveys with the SKAO, highlighting the role they can play in advancing this research field with respect to the pre-SKAO era.  Alongside well-explored scenarios for wedding cake-like, tiered extragalactic surveys at specific frequencies, we also address the scope for complementary efforts to obtain deep multi-frequency imaging over parts of (an) extragalactic field(s). In addition to providing key information on the physical properties of the emitting sources, such multi-frequency imaging will make important contributions to the calibration of observables from surveys with sparser radio spectral coverage. In this context, we explore possible pathways that can fully exploit the SKAO from initial (AA$^*$) to baseline capabilities (AA4). Finally, we highlight observational synergies with other major facilities -- for wide field and targeted follow-up science -- that will be operational in the 2030s, and for which joint coverage of extragalactic fields will generate significant legacy value.}

\begin{document}
\include{journal-names}

\maketitle

\section{Introduction}

Despite remarkable progress in recent years, fundamental questions remain unanswered in the field of galaxy formation and evolution.  We still lack a complete understanding of {\it when and how the first galaxies and their central black holes formed}, as well as of what physical processes governed the earliest episodes of star formation and metal enrichment. The {\it conversion of cosmic gas into stars} -- and especially the regulation of this process by feedback from supernovae and active galactic nuclei (AGN) -- remains a central uncertainty \citep{Tacconi2020}, as does the {\it role of environment and dark matter} (DM) halo occupancy in shaping the diversity of galaxies across cosmic time \citep{Wechsler2018}. The complex interplay between gas, stars, dust and DM halos -- and its dependence on cosmic large scale structure -- continues to challenge theoretical models \citep[e.g.,][]{Shuntov2022,Zhang2024}. Addressing these questions requires coordinated observations connecting the physical state of galaxies to their large-scale cosmological context, through the acquisition of statistically robust, unbiased, and well characterized samples. 

In this context, the SKA Observatory (SKAO) will play a pivotal role because of its unparalleled survey speed. Beyond its unprecedented ability to map neutral hydrogen (HI) throughout cosmic time, SKA extragalactic continuum surveys will transform our understanding of star formation (SF) and AGN activity. With exquisite sensitivity and dynamic range, as well as multi-frequency and multi-resolution capabilities, these surveys will deliver dust-unbiased, spatially resolved measurements of star-formation rates, magnetic fields, and cosmic-ray processes, providing a unique window into the energetics and feedback mechanisms that regulate galaxy growth. They will trace the radio emission from billions of galaxies, from the nearby Universe to the Epoch of Reionization (EoR), as well as jetted AGN down to the lowest radio powers, offering a statistically robust view of how SF and black hole accretion co-evolve over cosmic time, and providing unique insights into the role of jet-induced AGN feedback. 

Nevertheless, SKA surveys will need to be complemented by observations at shorter wavelengths to fully unlock their scientific potential. A panchromatic approach is essential for a comprehensive understanding of the complex process of galaxy formation and evolution, and of the concurrent growth of  super-massive black holes (SMBH) at galaxy centers. Only through observations across the full electromagnetic spectrum is it possible to get a full census of the physical (thermal and non-thermal) processes regulating star formation and nuclear activities in galaxies \citep{taba_18,Hassani,taba_24,Mazoochi}, as well as of the various galaxy components (stars, multi-phase gas, dust, and relativistic plasma), and link these to the evolving properties of galaxies as a whole \citep[see e.g.,][for an overview on synergies between SKAO and ESO facilities]{Prandoni2024}.

This chapter explores the broad subject of galaxy and AGN co-evolution, which represents a key scientific objective for extragalactic radio continuum surveys with the SKAO. We critically assess the scientific and technical requirements these surveys need to fulfill to unlock their transformational potential, emphasizing the need to implement  multi-scale, multi-tier, and multi-frequency strategies.
We also highlight key synergies with other facilities that will be operational in the 2030s. 

\begin{figure}[th!]
    \centering
	\includegraphics[width=0.99\columnwidth]{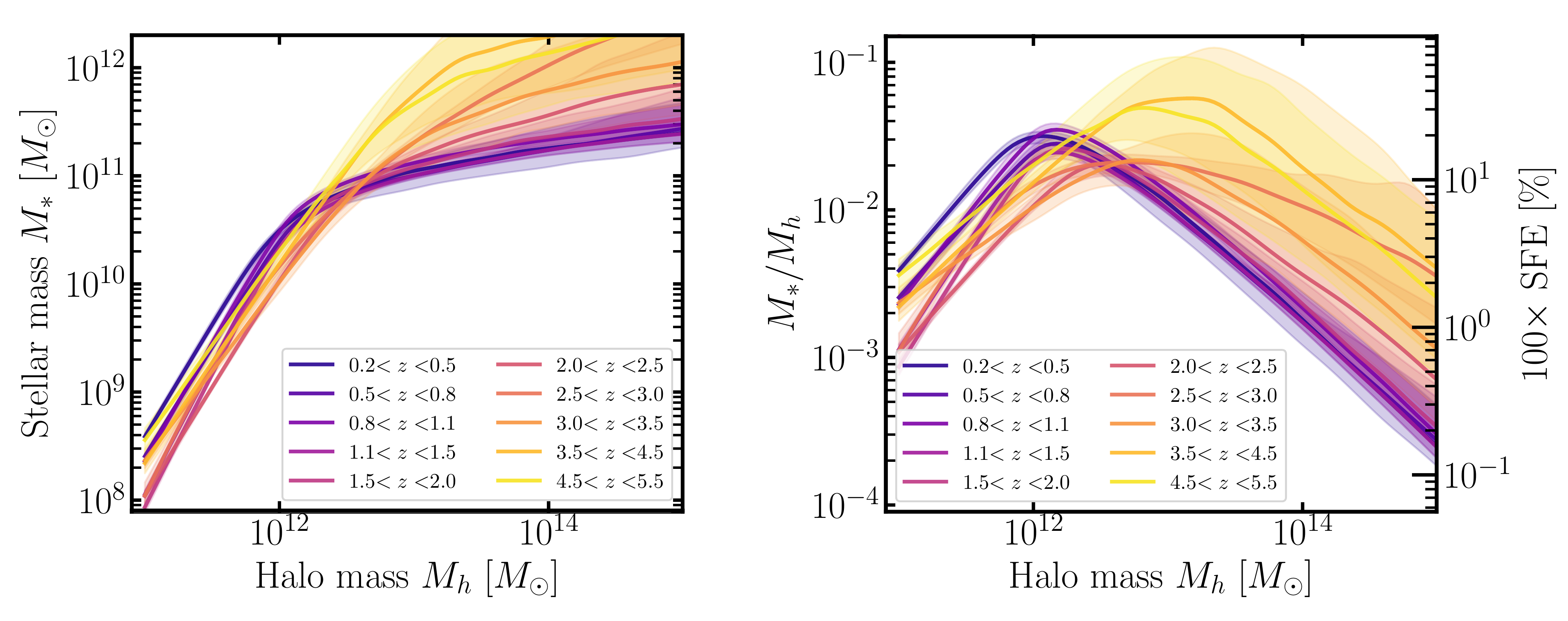}
    \caption{Stellar-to-halo mass relation (left) and $M_{\star}/M_h$ ratio, a proxy of star formation efficiency (right), in ten redshift bins, as measured in the COSMOS field. The shaded regions indicate 1$\sigma$ confidence intervals. Figure adapted from \citet{Shuntov2022}.}
    \label{fig:shmr}
\end{figure}

\section{Exploring the SMBH-galaxy-halo connection across cosmic time}\label{subsec:shmr}

In the Lambda Cold Dark Matter ($\Lambda$CDM) framework, galaxies emerge at the centre of dark matter halos \citep{Silk2012,Somerville2015},
accreting gas and forming stars in close inter-connection with the growth path of their host halos. This so-called galaxy–halo connection 
can be investigated through statistical relationships that connect galaxies to their environments and/or DM halo occupancy, such as the stellar-to-halo mass relation  \citep[SHMR;][Fig.~\ref{fig:shmr}, left panel]{Wechsler2018}.
On the micro-scale, SMBHs accrete gas and grow tightly connected to properties of the host galaxies \citep[such as the BH–galaxy/group scaling relations,][]{Kormendy2013,Gaspari2019}.
A number of studies have reported an increasing incidence of jetted AGN in massive galaxies at $z\lesssim 0.3$ \citep{Heckman2014}, which reaches close to 100\% at stellar masses $M_*>10^{11.5}$ M$_{\odot}$ \citep{Sabater2019}. This seems to indicate that the host mass plays a major role in driving the formation and evolution of a radio AGN (i.e. their life cycle). Cosmic environment can also play a role, as radio AGN are frequently associated with (passive) galaxies at the center of clusters and groups at low/intermediate redshifts \citep[$z\lesssim 0.5$;][]{Best2007}. Energetic SMBH-related processes within galaxies may impact their surroundings, at galactic and/or halo scales, influencing future gas accretion and star formation. Feedback from radio-loud AGN, in particular, is often invoked to explain the observed properties of massive galaxies in the local Universe:
radio jets are indeed thought to be responsible for the observed suppression of SF in massive, central galaxies (for DM halo masses $M_h\,{>}\,10^{12}\,M_{\odot}$; see Fig.~\ref{fig:shmr}, right panel).
Less clear is the role of jet-induced feedback at higher redshifts ($z \geq 1$), where radio-AGN activity shifts towards lower-mass and mostly star-forming galaxies \citep[SFGs;][]{Magliocchetti2022}, while the peak of the star formation efficiency (SFE) may move toward more massive halos (Fig.~\ref{fig:shmr}, right). 
Given that different forms of AGN feedback are implemented in current semi-analytic and hydrodynamic simulations of galaxy formation \citep{Bower2006,Croton2006,Hopkins2012},
constraining the AGN feedback modes, occurrence, and power as a function of environment represents a fundamental test of such models. One key element is to understand the environment in which the different AGN populations reside across cosmic time, and to link this with the underlying DM distribution, allowing a more direct comparison with galaxy formation models and simulations. 

Deep, large-scale galaxy/AGN surveys can play a key role in constraining galaxy-halo connection models and shedding light on the mutual interplay between large scale environment, galaxy/SMBH growth and AGN feedback. To shed light on galaxy/SMBH co-evolution, observations need to probe a sufficient cosmological volume to include the rarest, most powerful AGN at the epoch when both SF and nuclear activities peaked ($1<z<3$; aka ``cosmic noon'') and beyond, as well as the most extreme environments. On the other hand, they should also be sensitive to low-power radio AGN populations, as well as to inactive (i.e. non-AGN) galaxies across the entire stellar and halo mass regime where AGN feedback is thought to be relevant ($M_h\,{>}\,10^{12}\,M_{\odot}$, or equivalently $M_{\star}\,{\gtrsim}\,5{\times}10^{10}\,M_{\odot}$ -- cf. left panel of Fig.~\ref{fig:shmr} -- i.e. larger than the so-called {\it characteristic mass} $M^*_{\rm SFGs}$ at the knee of the stellar mass function). The latter is essential to facilitate comparison with matched control samples of AGN hosts and non-AGN galaxies. One complication is that by cosmic noon around 85\% of the total SFR density of the Universe is dust-enshrouded \citep{Dunlop2017}. Similarly, the fraction of heavily obscured AGN grows from 10-20\% in the local Universe to 80-90\% at $z\sim 4$ \citep{Vito2018}. Radio continuum surveys, by virtue of being insensitive to dust/gas obscuration, thus represent a uniquely valuable tool for obtaining a complete and unbiased census of SFG \citep[e.g.,][]{An2026.SKA, Algera2026.SKA} and radio AGN over cosmic time \citep[][and see also \citealt{Kondapally2026.SKA}]{Mazzolari2024}. The resolving power of interferometric radio observations can play a particularly important role in addressing these questions, as it prevents source confusion even in dense structures like galaxy groups and (proto-)clusters. Deep SKA surveys probing a broad range of environments will thus be able to overcome currently inconclusive findings on the evolution of signature galaxy scaling relations such as the galaxy main sequence (MS), for which both environmental variations and independence on environment have been reported in the literature \citep[e.g.,][]{erfanianfar16, duivenvoorden16, leslie20}. Furthermore, the large concentration of dusty star-forming galaxies in high-$z$ over-densities makes radio observations a powerful tool for pinpointing such structures in the first place \citep[e.g.,][]{daddi17}.

As we will demonstrate in Sect.~\ref{sect:B2tieredsurv}, the wide and deep tiers of the Band 2 surveys defined by \cite{Prandoni2015} provide an excellent framework for investigations of the SMBH-galaxy-halo connection. Multi-tier Band 5 surveys and/or multi-frequency radio observations as those described in Sect.~\ref{sec:multif}, can also play an important role. They can be used to observationally constrain the age of the radio jet plasma, and derive radio AGN duty cycles (see also \citealt{Hardcastle2026.SKA}); they can provide diagnostics to identify radio AGN at the highest redshifts (\citealt{Afonso2026.SKA}), and/or AGN radio cores embedded in SF disks (\citealt{Panessa2026.SKA}). Similarly to AGN life-cycles, for SFGs, analysis and decomposition of radio spectral energy distributions (SEDs) in MHz-to-GHz multi-band imaging provide insights into the age of individual starburst episodes \citep[e.g.][]{Thomson19}, allows us to probe the SF activity on different time scales via free-free and synchrotron emission \citep[e.g.,][]{Moldon2026.SKA}, and more generally to study how thermal and non-thermal emission processes shape the energetics of the interstellar medium \citep[e.g.,][]{Tabatabaei2026.SKA}. Investigating radio SEDs as a function of redshift can also provide indications on k-correction evolutionary trends to be accounted for to obtain precise radio luminosities \citep{taba_25}. It is, however, important to emphasize that complementary data from survey facilities operating at other wavelengths are essential for a full characterization of the physical properties of the radio sources (e.g. redshift, stellar mass, monochromatic and/or bolometric luminosities, etc.), and for tracing the AGN fueling and feedback cycle (\citealt{Maccagni2026.SKA}). Such multi-wavelength observations are also crucial for mapping the large-scale cosmic structure and the DM halos in which these sources reside. Such studies thus are an excellent example of the value of synergistic, multi-facility approaches (see Sect.~\ref{sec:synergies} for more details).

\section{Survey strategies}

In this section, we connect the key scientific questions we aim to address with the SKAO to the corresponding survey requirements. Forecasts will be based on the T-RECS simulated radio catalogues \citep[][and references therein]{bonaldi23}. We also assess whether the suite of extragalactic continuum {\it reference} surveys defined by \citet{Prandoni2015} to address galaxy/AGN co-evolution with the SKAO 
remains competitive in light of the legacy surveys conducted with SKAO precursors and pathfinders over the past decade, and whether they are suited to address the science outlined in Sect.~\ref{subsec:shmr}.  

\subsection{A multi-tier survey for galaxy/SMBH co-evolution studies with SKA-Mid Band 2}
\label{sect:B2tieredsurv}

\subsubsection{Source populations probed by different survey tiers}
\label{sect:tierpops}

\begin{figure}[t!]
    \centering
	\includegraphics[width=0.8\columnwidth]{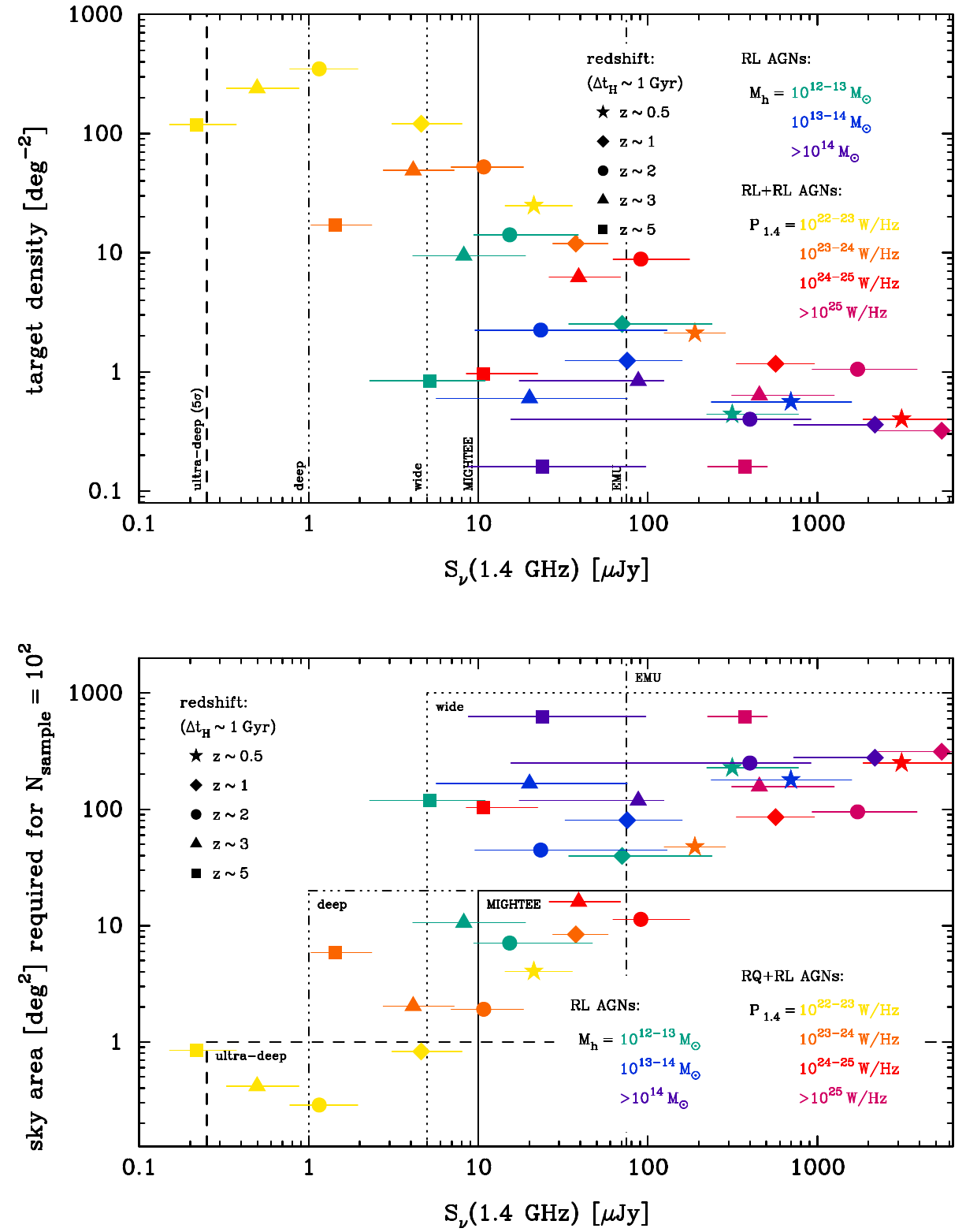}
    \caption{Upper panel: Sky density of AGN populations in the T-RECS simulation \citep{bonaldi23}, as a function of their predicted median flux density at 1.4\,GHz. Horizontal error bars span the interquartile range (25-75\%) of the fluxes tabulated in the T-RECS catalog for a given AGN sub-sample. AGN populations are binned by redshift (and drawn from a $\sim$1\,Gyr interval centred on the reference redshifts listed in the figure legend). Radio-loud (RL) AGN with $P_{1.4}\,{>}\,10^{23}$\,W/Hz are additionally split by host dark matter halo mass (green/blue/indigo symbols). The combined population of radio-quiet (RQ) and RL AGNs is split by luminosity (yellow/orange/red symbols). The selection of RQ and RL objects is described in Sect. \ref{sect:tierpops}. Vertical lines are drawn at the 5$\sigma$ sensitivities of the extragalactic continuum reference survey tiers (ultra-deep/deep/wide) outlined in Table~\ref{tab:2015sur}, and of the precursor surveys MeerKAT/MIGHTEE \citep{Jarvis2016} and ASKAP/EMU \citep{Norris2011,Norris2021,Hopkins2025}.\newline
    Lower panel: Sky area required to reach a sample size of at least 100 objects for all different AGN sub-samples vs. their predicted 1.4\,GHz median flux density and 25-75\% flux density spread. All surveys indicated probe the parameter space below (to the right of) the horizontal (vertical) lines in the figure.}
    \label{fig:AGNs_in_tiers}
\end{figure}

\begin{figure}[th!]
    \centering
	\includegraphics[width=0.8\columnwidth]{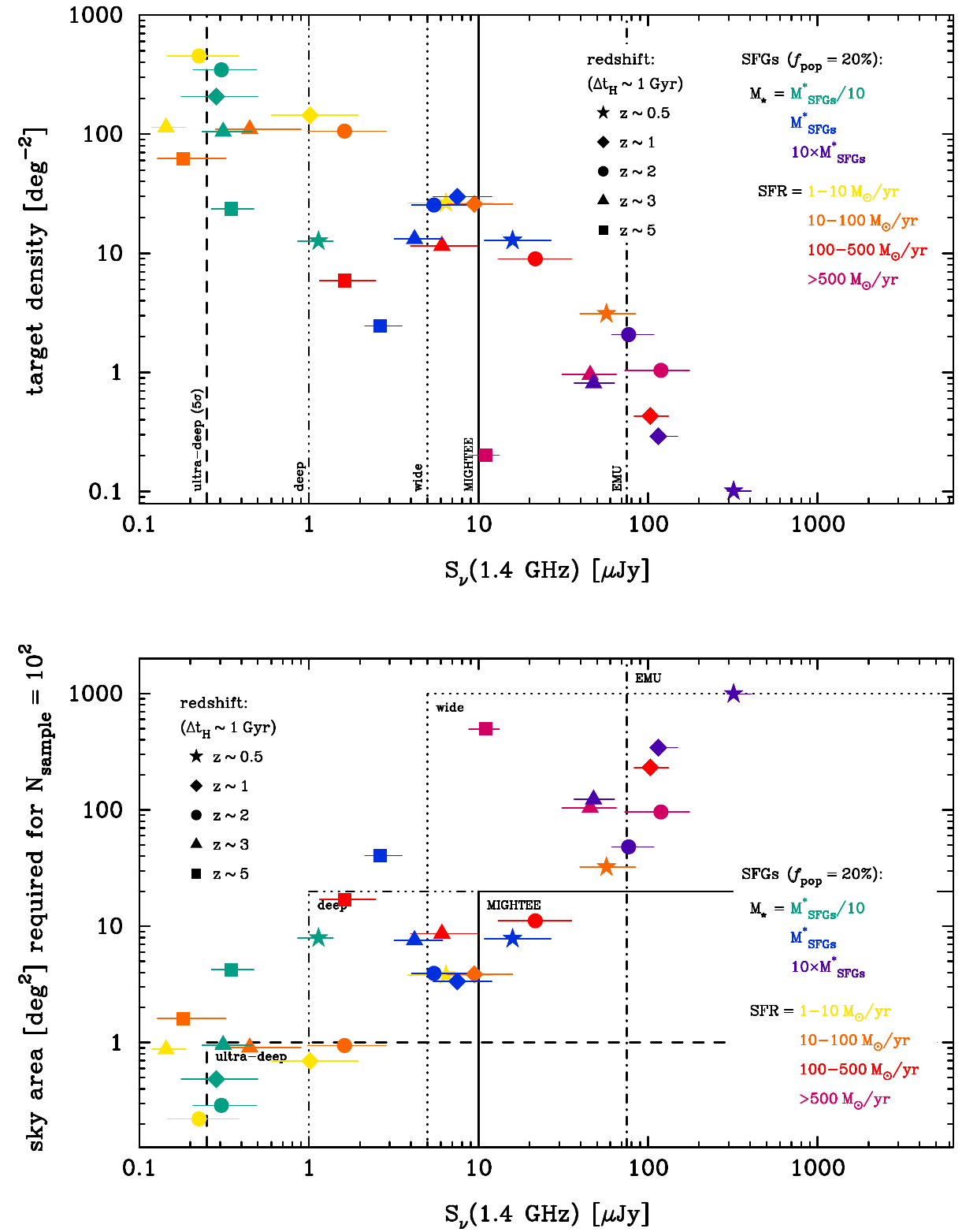}
    \caption{As in Fig. \ref{fig:AGNs_in_tiers}, but focusing on sub-samples of the star-forming galaxy (SFG) population. SFGs are split by (i) stellar mass (green/blue/indigo symbols), relative to the characteristic mass $M^*_{\rm SFGs}$ at the knee of their stellar mass function, and (ii) galaxy SFR (yellow/orange/red symbols). All predicted number densities have been divided by a factor 5 ($f_{\rm pop}$\,=\,20\%), to facilitate comparative studies among SFGs with given set of physical properties, e.g., as a function of environment or SF history (see Sect. \ref{sect:tierpops}).}
    \label{fig:SFGs_in_tiers}
\end{figure}

Given the shape of galaxy and AGN luminosity functions, a wedding cake, multi-tier survey design  ensures that both (i) intrinsically bright but rare source populations, as well as (ii) faint but abundant targets can be probed in representative numbers. We use the T-RECS simulated catalogues to locate where different galaxy populations reside in the parameter space covered by the tiered Band 2 surveys of \citet{Prandoni2015}, also outlined in Table \ref{tab:2015sur}. Specifically, we focus on different samples of AGN hosts and SFGs, which we split by either radio brightness (directly linked to the intrinsic activity levels) or characteristic mass scales/environment, and trace these through the observational parameter space of the survey tiers as a function of redshift out to $z\,{\sim}$\,5.\\
In the T-RECS simulation, the ``radio-loud'' (RL) AGN population consists of a mixture of steep-spectrum sources, flat-spectrum radio quasars and BL Lacs, with different evolutionary properties based on the prescriptions in \citet{bonato17}. The model adopted in T-RECS for the luminosity functions of these three flavours of RL AGN implies that they extend in luminosity well below the radio power threshold of $P_{\rm1.4\,GHz}\,{\gtrsim}\,10^{23}$\,W/Hz, which is often adopted to separate  RL AGN from RQ AGN and SFG populations (\citealt{Magliocchetti17}).  The radio emission from RQ AGN (i.e. Seyfert galaxies and RQ-QSO) is assumed to be SF-driven in the T-RECS simulation. RQ AGN are consequently considered part of the  SFG population and are not modelled separately. However, multi-band diagnostics \citep{Delvecchio2017} and/or VLBI spatial resolution surveys \citep{muxlow20,Radcliffe2021a,Radcliffe2021b} have shown that AGN-driven radio emission can also be present in $\sim$30\% of RQ AGNs. To provide estimates of AGN number densities and characteristic flux ranges from the RQ into the RL luminosity regime, we make the simplifying assumption in this chapter that RQ AGNs constitute a constant 30\% of the SFG population at all fluxes. Radio sources are associated to DM haloes via a clustering model in T-RECS. In Fig. \ref{fig:AGNs_in_tiers} we exploit the DM halo mass information in the T-RECS catalog to estimate the locus of radio AGN in massive DM halos ($M_h\,{\geq}\,10^{12}\,M_{\odot}$) in the parameter space of a tiered survey. Acquiring a sufficient number of radio AGN at these halo mass scales is indispensable for reconstructing how AGN feedback leads to the reduced baryon-over-DM mass ratios inferred for the most massive DM halos (see Fig. \ref{fig:shmr}, right).\\
For SFGs, a step-change in our understanding will flow from being able to probe SF activity in a dust-free manner for galaxy samples large enough that multi-variate analyses become routinely possible, e.g., by fixing a specific set of physical properties and exploring the origins of the scatter of the SF population with respect to another (e.g., galaxy environment, offset from scaling relations, or galaxy classifications based on structure/morphology or emission line diagnostics). In Fig. \ref{fig:SFGs_in_tiers} we account for this need to slice and dice the SF population in a variety of ways by dividing the number densities of the corresponding sub-samples of SFGs by a factor 5 ($f_{\rm pop}$\,=\,20\%), corresponding to -- at fixed redshift and stellar mass or SFR -- 5 additional bins with respect to a third galaxy property. For SFGs, in addition to binning by redshift, we consider two further key properties in Fig. \ref{fig:SFGs_in_tiers}: (i) SFR, spanning the range from galaxies with Milky Way-like activity levels at 1--10\,$M_{\odot}$/yr to the ULIRG/HyLIRG-regime, and (ii) stellar mass ($M_{\star}$). The binning in stellar mass is relative to the characteristic mass $M^*_{\rm SFGs}$ at the knee of the stellar mass function of SFGs, with $M^*_{\rm SFGs}$ remaining roughly constant at ${\sim}5{\times}10^{10}\,M_{\odot}$ over the redshift range we explore here \citep[e.g.][]{weaver23}. 

\begin{table}
\footnotesize
\centering
  \caption[Landscape table]{Outline of multi-tier SKA-Mid {\it reference} surveys (adapted from \citealt{Prandoni2015}) with observing time estimates.}  
  \label{tab:2015sur}
 \begin{tabular}{clccc|ccc|ccc}
 \hline
 Band &  Tier & rms & Area & $\theta$ & $\nu_{\rm ref}$ & BW & A$_{eff}$  & Briggs / $\theta$ & t/point  &   t/deg$^2$\Tstrut\\ 
   &  &  [$\mu$Jy/bm] & [deg$^2$] & [arcsec] & [GHz] & [GHz] & [deg$^2$] & / [arcsec] & [hr]  & [hr]\\[1ex]
        \multicolumn{1}{c}{(1)} & \multicolumn{1}{c}{(2)} & (3) & (4) & (5) & (6) & (7) & (8) & (9) & (10) & (11)\Bstrut\\
   \hline
 \multicolumn{1}{c}{2} & \multicolumn{1}{c}{\parbox{1cm}{Ultra- Deep}} & 0.05 & $1$ & $0.5$ &  1.35 & 0.34 & 0.55 & -1 / 0.45 & 2667  &  4849\Ttwostrut\\[2.5ex]
           & \multicolumn{1}{l}{Deep} & 0.2 & 10-30 &  0.5   & & & & & 167 & 304\\[1.5ex]
            & Wide & 1 & \multicolumn{1}{c}{1000} & 0.5 &  & & & & 6.7 & 12.2 \Bstrut\\
\hline
  \multicolumn{1}{c}{5b}  & \multicolumn{1}{c}{\parbox{1cm}{Ultra-Deep}} & 0.04 & 0.04 & $\lesssim 0.1$  & 11.85 & 4.2 & 0.007 & 0 / 0.08 & 275 & 3146$^*$\Ttwostrut\\[2.5ex]
    & Deep & 0.2 & 2 &  $\lesssim 0.1$   &    &  &  &  & 11 & 1573\Bstrut\\
\hline\\[-2ex]
  \end{tabular}
\begin{minipage}{\textwidth}
    \vspace{0.2cm}
    \small
    {\it Notes:} The Bandwidth (BW) settings in column 7 take into account RFI occupancy in the band (see Sect.~\ref{sec:multif2}). The noise effective area in column 8 is the area over which the sensitivity is uniform, and is defined as $A_{eff}=2340\,(\lambda_{ref}/D)^2$, with $D=15$\,m. Exposure times in column 10 are calculated with the SKAO Sensitivity Calculator assuming Decl. $=-45^{\circ}$, Elevation $=45^{\circ}$. In addition, for Band 5b exposure time and resolution estimates we assume also MeerKAT antennas are equipped with Band 5b receivers; this results in 1.15$\times$ coarser resolution and 0.52$\times$ shorter integration times (see Sect.~\ref{sec:multif2} for details). Finally we note that the ultra-deep sensitivity requirement (see column 3) can only be reached at ${\lesssim}1.5''$ angular resolutions to mitigate confusion noise (see also Fig.~\ref{fig:time_FWHM}). $^*$ This time refers to the survey area quoted in column (4), i.e. 0.04 deg$^2$.  
    \end{minipage}
\end{table}

\begin{figure}[t]
    \centering
    \vspace{-2cm}
	\includegraphics[width=0.97\columnwidth]{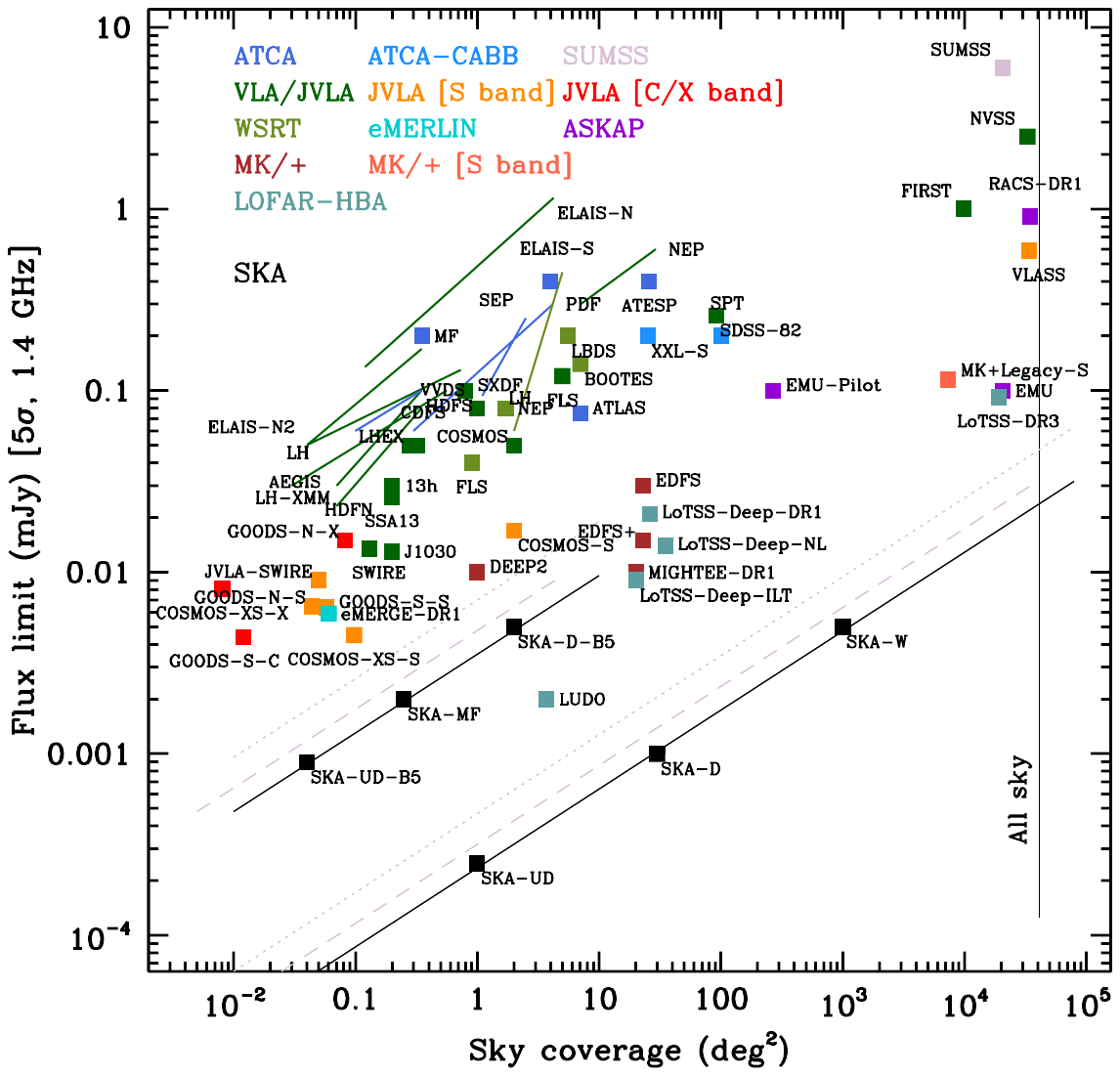}
    \vspace{-3.5cm}
    \caption{SKA-Mid Band 2 and Band 5 reference surveys, as defined in Table \ref{tab:2015sur}, in comparison with existing surveys: area coverage vs depth (5$\sigma$ flux limit). Also shown is the deep SKA multi-frequency survey (SKA-MF) optimized for SFG science, discussed in Sect. \ref{sec:multif2}. Flux limits are rescaled to 1.4\,GHz (assuming $\nu^{-0.7}$) where necessary. Different colors highlight surveys undertaken with different facilities (see legend). The pale violet diagonal lines show the parameter space probed by SKA surveys 2$\times$ smaller (dashed lines) or 2$\times$ shallower (dotted lines) than those illustrated in Table \ref{tab:2015sur}. 
    }
  \label{fig:SKAsurveys}
\end{figure}

As shown in Fig. \ref{fig:AGNs_in_tiers} (top) a complete census of RL AGNs, over a wide range of radio powers (P$_{\rm 1.4\,GHz}\,{>}\,10^{23}$\,W/Hz), redshifts ($0\,{<}\,z\,{<}\,5$) and environments (up to $M_h\,{>}\,10^{14}\,M_{\odot}$), requires radio surveys with a 5$\sigma$ flux limit at $S_{\nu}({\rm 1.4\,GHz})\,{\sim}$\,1\,$\mu$Jy (`Deep' tier). Similar sensitivities can fully probe inactive galaxies down to the characteristic stellar mass $M^*_{\rm SFGs}$ (Fig. \ref{fig:SFGs_in_tiers}, top). The additional need to achieve robust statistical sampling of the diverse properties of radio AGN and mass-matched inactive galaxies necessitates multi-tiered radio surveys covering areas from 10s to 1000 square degrees (Figs. \ref{fig:AGNs_in_tiers} and \ref{fig:SFGs_in_tiers}, bottom panels). We note that while wide-area coverage can be obtained by combining a number of smaller surveys, helping to beat cosmic variance, a single survey spanning a large connected area would be preferable for galaxy-halo connection studies (see Sect.~\ref{subsec:shmr}), as it would imply better statistics for the two- (three-)point correlation functions, ensuring statistically robust, multi-variate, clustering studies of the various galaxy/AGN populations. Sampling wide areas also allows us to better probe the regime where the growth of perturbations is still linear. On the other hand, the ultra-deep tier ($S_{\nu}({\rm 1.4\,GHz})\,{\gtrsim}$\,0.25\,$\mu$Jy) is essential for obtaining a full census of SFGs and/or RQ AGNs, and provides robust statistical samples down to the lowest stellar masses and up to the highest redshifts (see Figs.~\ref{fig:AGNs_in_tiers} and~\ref{fig:SFGs_in_tiers}). Also interesting is the comparison with pre-SKAO surveys.  Fig. \ref{fig:AGNs_in_tiers} shows that EMU plays a unique role for tracing the most luminous AGN ($P_{\rm1.4\,GHz}\,{>}\,10^{25}$\,W/Hz) and extreme environments ($M_h\,{\simeq}\,10^{14}\,M_{\odot}$), but only the SKA deep and wide tiers can fully sample fainter RL AGN and lower mass DM halos. On the other hand, MIGHTEE can fully sample radio AGN up to redshifts $z\,{\sim}\,2$, but only the SKA deep tier can push their study to higher redshift. Similarly, Fig.~\ref{fig:SFGs_in_tiers} clearly shows that the SKA deep and ultra-deep tiers play a key role to fully address SFG and RQ AGN populations, as EMU \citep{Norris2011,Norris2021,Hopkins2025}, MIGHTEE \citep{Jarvis2016} and superMIGHTEE \citep{Lal2025} can only trace bright and/or massive systems.

An overall comparison between the \citet{Prandoni2015} tiered Band 2 continuum surveys (with specifications as per Table \ref{tab:2015sur}) and other existing radio surveys is shown in Fig.~\ref{fig:SKAsurveys}. It is clear that the multi-tier SKA surveys defined back in 2015 remain competitive ten years later, still representing a significant step forward with respect to the legacy surveys conducted by SKAO precursors and pathfinders. It is worth highlighting that the SKA deep and ultra-deep tiers will significantly out-perform  also the ultra-deep survey planned for LOFAR 2.0. The orange and pale violet dashed lines in Fig.~\ref{fig:SKAsurveys} respectively show the parameter space probed by SKA surveys when relaxing the area and sensitivity constraints by a factor of 2. It is clear that these surveys remain competitive with respect to pre-SKA surveys. While more limited in science scope, such surveys may be better suited for SKA early operations (see  Sect.~\ref{sect:pathAAstar} for a discussion of AA$^*$ capabilities).

\subsubsection{Frequency and angular resolution requirements}\label{sec:angres}
For galaxy evolution and AGN studies we are mostly frequency agnostic as long as the requisite source density, galaxy star-formation rates, AGN radio powers are met.  Assuming $S\sim \nu^{\alpha}$, our quoted sensitivities (see Table~\ref{tab:2015sur}) should be simply scaled by the average spectral index of the dominant extragalactic source populations at the flux densities of interest to get corresponding sensitivities at other frequencies. For sources dominated by non thermal optically-thin synchrotron emission, like star-forming galaxies and radio jets, we have $\nu^{-0.7}$. Hence, as long as the required resolution is retained (see below), continuum surveys are generally better done at as low frequency as feasible. For SKA-Mid observing down to a frequency of ${\sim}600-700$\,MHz (Band 1) could be advantageous, due to the larger field-of-view, and hence survey speed, at lower frequencies. Observing in SKA-Mid Band 1 would also exploit commensality with cosmological HI surveys, and provide HI line information (HI masses/redshifts) for at least some classes of galaxies. Observations at a higher frequency (e.g. $\sim$1.4\,GHz), on the other hand, would exploit commensality with polarization surveys (magnetism) and would naturally provide finer angular resolution, an important requirement for galaxy evolution studies.

\begin{figure}[t]
    \centering
	\includegraphics[width=0.7\columnwidth]{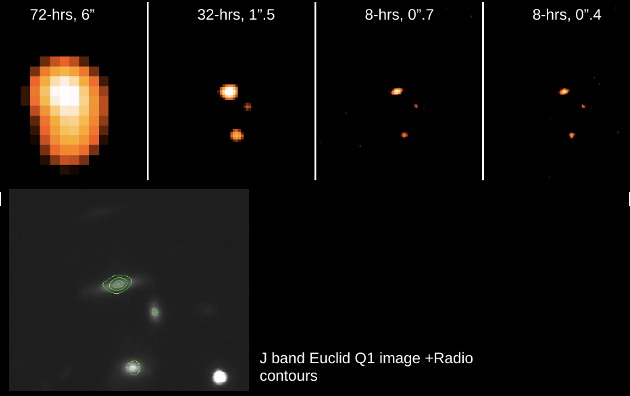}
    \vspace{0.5cm}
    \caption{{\it Top row:} LOFAR images obtained at various angular resolutions and integration times. From left to right: resolution improvement from the standard 6$''$ provided by NL baselines, to 1.5, 0.7 and 0.4$''$ when including international stations, with different weightings. With international stations one can de-blend the source into three different components. {\it Bottom:} J-band Euclid image from the Quick Release 1 (Q1; \citealt{Ausssel2025}) with radio contours overlaid, showing that the three radio components are in fact three distinct radio sources, each with their own optical counterpart. (M. Bondi, priv. comm.)}
    \label{fig:EDFNmultires}
\end{figure}

Angular resolution is a critical parameter for galaxy evolution studies for several reasons, the most obvious being avoidance of source confusion. The Band 2 ultra-deep and deep tiers outlined in Table~\ref{tab:2015sur} would be confusion-limited if undertaken at, respectively, ${\gtrsim}1.5''$ and ${\gtrsim}2.5''$ resolution (SKAO Sensitivity Calculator). Another critical aspect is the ability to reliably identify and classify radio sources at any redshift. This requires extensive, deep multi-wavelength information in the surveyed areas, as well as sub-arcsec resolution radio observations, as clearly illustrated in Fig~\ref{fig:EDFNmultires}. The resolution requirement of 0.5$''$ indicated in Table~\ref{tab:2015sur} for Band 2 tiers provides a good match with the resolution of state-of-the-art optical/NIR surveys (0.3$''$ -- {\it Euclid}; 0.7$''$ -- LSST; see Sect.~\ref{sec:synergies} for synergies between the SKAO and survey facilities operating at other wave-bands). As discussed in Sect.~\ref{sec:b5res}, ultra-high-resolution (e.g. ${\lesssim}0.1''$),  and even VLBI resolution, are essential to morphologically separate compact AGN and radio cores from extended, SF-related emission (see \citealt{Baldi2026.SKA,Panessa2026.SKA}). 

\subsection{The importance of multi-frequency imaging for galaxy-SMBH co-evolution science}\label{sec:multif}

\subsubsection{Exploiting the angular resolution of SKA-Mid Band 5 for resolved SFG and AGN studies}\label{sec:b5res}

As extensively discussed by \cite{Murphy15}, we gain significantly by having matched-sensitivity Band 5 multi-tier surveys.  

Galaxies are found to shrink to $\lesssim  5$ kpc sizes at redshifts $z\gtrsim 1$ (see Fig.~8 of \citealt{Costantin2023} for galaxies with $M_{\star}\,{>}\,10^9\,M_{\odot}$), corresponding to ${\lesssim}0.6''$ scales. So radio observations of similar or better resolution are ideally needed to reliably investigate galaxies at cosmic noon and beyond (see also bottom panel of Fig.~\ref{fig:EDFNmultires}). The recoverability of the structural properties of marginally resolved galaxies in $\sim$0.5$''$ SKA maps has been discussed by \citet{coogan23}, with surveys like e-MERGE \citep{muxlow20} providing early glimpses into the spatial details present in distant galaxies imaged at GHz frequencies on even smaller scales of $\sim$0.1$''$. This $\sim$10--100-fold better resolution than the arcscond scale would be needed to perform resolved studies of SF processes in high-$z$ galaxies. Band 5b surveys can play a critical role in this respect thanks to their ability to provide imaging at 0.05-0.1$''$ at $\sim10$ GHz. Additionally, Band 5b probes different emission processes (thermal vs. non-thermal) at different redshifts: for galaxies at $z\sim 2-3$, observations at $\sim$10\,GHz sample a rest-frame frequency of 30--40\,GHz, i.e. the spectral regime where free-free begins to dominate over synchrotron emission (see also Sect. \ref{sec:multif2}). 

The discovery of an increasing fraction of radio sources associated with RQ AGNs in the current deepest radio fields (\citealt{Delvecchio2017,Whittam2022,Best2023}) means that sensitive radio surveys can in principle probe the entire AGN population, and not only the tiny RL AGN fraction (${\sim}10-20\%$; \citealt{Urry95,Kratzer15,Macfarlane21}). Galaxy-AGN co-evolution studies would therefore significantly benefit from sensitive tiers in Band 5 (see Table~\ref{tab:2015sur}), which would allow us to pinpoint flat-spectrum radio cores at the center of SFGs. In this respect, the availability of VLBI capabilities for the SKAO can make key contributions as well (see \citealt{Panessa2026.SKA})\footnote{Another way to identify AGN cores in galaxies is through variability studies (see Sect.~\ref{sect:variability_scdrivers} for more details).}. Morphological separation of centrally concentrated and extended SF provides fundamental constraints on the balance between AGN and SF activity (key especially for understanding radio emission from radio-quiet QSOs), and for pure SF systems radio size measurements provide insight into the growth mode of galaxies - e.g., disk-wide, secular growth of MS galaxies vs. concentrated, possibly interaction-driven starbursts (i.e. systems with high specific star formation rates; \citealp{jimenez-andrade19}).  As dust obscuration is subject to strong spatial variations (with central regions being particularly dust-attenuated in UV/optical imaging, e.g., \citealp{morselli19, nelson19}), dust-unbiased, high-resolution radio maps hold a particular potential for directly linking the overall structural evolution of galaxies (e.g., disk growth and bulge formation) to the temporal and spatial variations/evolution of SF and feedback processes for studying the build-up of different galaxy components.

However, because of the significantly smaller field of view in Band 5, it will not be possible in all flux density regimes to cover the full area of, e.g., surveys in Band 2. In Table~\ref{tab:2015sur} we outline a two-tiered (ultra-deep and deep) Band 5b survey, that provides matched sensitivity to the deep and wide Band 2 tiers, albeit over a limited sky area. We note that this survey is similar to the one proposed by \citet{Prandoni2015}, with the only difference that we slightly adjusted the deep tier parameters: the rms sensitivity requirement decreased from 0.3 to 0.2\,$\mu$Jy/bm to better match the Band 2 wide tier sensitivity, while the sky coverage was doubled (from 1 to 2 deg$^2$). With this modification the SKA band 5 survey becomes more competitive with respect to the new high-frequency (S, C and X bands) surveys  carried out by the JVLA  over the past ten years (see orange and red filled squares in Fig.~\ref{fig:SKAsurveys}).

\subsubsection{A matched-resolution, multi-frequency deep field for galaxy-integrated radio SEDs }
\label{sec:multif2}

The versatility of the SKA telescopes (sensitivity across a wide range of angular scales, beamforming capabilities for LOW, and options for sub-array-mode observations in general) make it possible to pursue observing strategies aimed at matching the observational parameter space of multi-frequency projects in a variety of ways. In this section, we describe an extragalactic survey that is matched across multiple frequency bands in terms of angular resolution and spectral response, and -- to the extent this is possible -- aims to cover a somewhat commensurate sky area across all bands \citep[e.g., see also the superMIGHTEE project:][]{Lal2025}. Depending on the specific science goals, a multi-band project might prioritise matching the band-to-band survey area in a more strict sense, or instead opt for obtaining simultaneous observations at several frequencies by exploiting the sub-array capabilities of the SKAO. The latter could be particularly beneficial for characterising time-varying source populations in multi-epoch surveys (see Sect. \ref{sect:variability_scdrivers}).

We set the following requirements for our multi-frequency survey:
\begin{enumerate}
\item span $\sim$2\,dex in frequency (0.15--15\,GHz), with no more than 50\% variation in angular resolution (and beam sizes varying no more than 10\% between 300\,MHz and Band 5b),
\item ensure that thermal noise remains at least a factor 1.5 above the confusion noise in all bands (i.e. the confusion noise contributes $\lesssim$30\% to the total noise budget at the flux limit), and
\item achieve a sensitivity and sky area sufficient for a robust characterisation of the radio SEDs of (i) $M^*$ galaxies (${\sim}5{\times}10^{10}\,M_{\odot}$) throughout the peak era of the cosmic SFH (out to z\,$\sim$\,3), thereby jointly constraining the cosmic SFR density through both radio free-free and synchrotron emission, and (ii) galaxies with Milky Way-like SFRs (1--10\,$M_{\odot}$/yr) out to $z\,{\sim}\,1$.
\end{enumerate}
Throughout this section, we focus on observations with AA4. In Sect. \ref{sect:pathAAstar} we then briefly discuss the scope for implementing some components of such a multi-band survey already in AA$^*$.

To define a suitable observing strategy, we use the options in the SKAO Sensitivity Calculator\footnote{~\texttt{https://sensitivity-calculator.skao.int}} to apply different $uv$-plane weighting schemes and tapering scales when calculating the expected image noise. We adopt the default settings of the sensitivity calculator for target elevation, and assume a survey field centred at Dec.\,=\,-45$^{\circ}$. In defining tuning ranges in the sensitivity calculator, we aim for a fractional bandwidth of at least $\Delta\nu/\nu_{\rm ref}$\,=\,0.3 (with $\nu_{\rm ref}$ being the central frequency of the band), but where necessary reduce the instantaneous bandwidth $\Delta\nu$ to account for the RFI environment as currently characterised at the SKAO sites or other relevant facilities (e.g., the JVLA). Our choices are as follows (see also Table \ref{tab:multifreqobsparams}):
\begin{itemize}
\item SKA-Low: three bands centred at $\nu_{\rm ref}$\,=\,80, 160 and 300 MHz with instantaneous bandwidth $\Delta\nu$\,=\,26, 51 and 60\,MHz. For the 300\,MHz tuning achieving a fractional bandwidth $\Delta\nu/\nu_{\rm ref}$\,=\,0.3 is unfeasible due to persistent satellite RFI features between 243.5--270.4, 277.5--285.0 and 334--342\,MHz \citep{mckay25}.
\item SKA-Mid: tunings centred at 700 (B1), 1.355 (B2), 6.55 (B5a) and 11.85\,GHz (B5b). We note that for Band 5b, two offset spectral windows with $\Delta\nu\,{\sim}$\,2\,GHz at 9.5 and 14.1\,GHz will in practice be necessary to avoid RFI in the Starlink download window from 10.7--12.7\,GHz. The instantaneous bandwidths are 400\,MHz (B1\footnote{~For B1, in addition to RFI, the narrower 0.58--1.015\,GHz UHF band tuning range of the MeerKAT dishes compared to the SKA-Mid dishes is a further factor in determining the frequency range over which the sensitivity will be maximised.}), 340\,MHz (B2), 3.3\,GHz (B5a) and 4.2\,GHz (B5b), in keeping with RFI detected in current MeerKAT UHF and L-band data\footnote{~https://skaafrica.atlassian.net/wiki/spaces/ESDKB/pages/305332225/Radio+Frequency+Interference+RFI-UHF-Statistics} and in JVLA C-/X-band observations (fraction of bandpass impacted at the JVLA: $\sim$15\%\footnote{https://science.nrao.edu/facilities/vla/docs/manuals/obsguide/rfi}).  
\end{itemize}

\begin{table}[t!]
	\centering
	\caption{Depth-independent multi-frequency survey parameters.}
	\label{tab:multifreqobsparams}
	\begin{tabular}{ccccccc}
		\hline
		telescope  & array config. &  $\nu_{\rm ref}$ & $\Delta{\nu}$ & FHWM$_{\rm bm}$ & robust & taper\Tstrut\\
         &  & [GHz] & [GHz] & [arcsec] & & [arcsec]\\[1ex]
        \multicolumn{1}{c}{(1)} & (2) & (3) & (4) & (5) & (6) & (7)\Bstrut\\
		\hline
		SKA-Low & AA4 (outer 24\,km) & 0.080 & 0.026 & 5.14 & -1 & -- \Tstrut\\
        & AA4 (outer 12\,km) & 0.160 & 0.051 & 3.51 & -1 & -- \\
        & AA4 & 0.300 & 0.060 & 2.53 & -1.5 & -- \\[1.5ex]
        SKA-Mid & AA4 & 0.800 (B1) & 0.400 & 2.50 & 1 & 0.54 (`0.31') \\
        & AA4 & 1.355 (B2) & 0.340 & 2.50 & 1 & 0.58 (`0.56') \\
        & AA4 & 6.550 (B5a) & 3.300 & 2.45 & 1 & 0.86 (`4') \\
        & AA4 & 11.850 (B5b) & 4.200 & 2.40 & 1 & 0.81 (`6.82')\Bstrut\\
		\hline
	\end{tabular}
    \begin{minipage}{\textwidth}
    \vspace{0.2cm}
    \small
    {\it Notes:} $\nu_{ref}$ in Column 3 is the observational tuning frequency. Angular resolutions FWHM$_{\rm bm}$ in col. 5 are beam sizes after applying Briggs weighting with the robust parameter in col. 6, and a Gaussian taper with the angular scale listed in col. 7. Numbers in brackets in col. 7 give the corresponding value of the dimensionless tapering parameter used in the SKAO sensitivity calculator (see Sect. 2.2.1 in the SKA-Mid Sensitivity Calculator User Guide).
    \end{minipage}
\end{table}

We also introduce a couple of refinements with respect to default sensitivity calculator outputs. Firstly, we scale the Band 5b calculator outputs to account for the additional receivers deployed by INAF/MPG on all MeerKAT dishes. Using a simple scaling based on the relative change of collecting area, this improves sensitivity by a factor 1.39 for AA4, (reducing B5b observing times by a factor 0.52 as a consequence). As MeerKAT dishes are centrally concentrated within the SKA-Mid array, the angular resolution becomes coarser by 15\%\footnote{~The change in angular resolution is somewhat dependent on the weighting scheme adopted. The 15\% quoted are a representative average of sensitivity calculator results for AA4 with Briggs weighting and robust parameters between -2 and 1. For AA$^*$ the angular resolution change amounts to 20--25\% if not only 15\,m SKA-Mid dishes, but also MeerKAT dishes are included in the array.}. Reflecting this small change in angular resolution, we have also increased the confusion noise level by a factor 1.32, based on Fig. 10 in SKA-Mid Sensitivity Calculator User Guide.\\
Our second refinement addresses trade-offs between angular resolution and image noise. A comparable image resolution can only be achieved across our full frequency range if next-to-uniform weighting is adopted for SKA-Low\footnote{~The angular resolution of SKA-Low imaging can be further maximised using the ``outer \textit{X}\,km" configuration, at the cost of reduced sensitivity. The sensitivity loss can be partly compensated by adopting a robust parameter of -1 rather than -2. For $X\,{\in}\,\{12, 24\}$, given the $uv$-space sampling of the SKA-Low stations involved, this change in robust parameter produces a substantially smaller difference in synthesized beam size ($\lesssim$2\%) than it would for the full SKA-Low array.}, and a weighting scheme approaching `natural' in SKA-Mid bands\footnote{~Natural weighting (robust\,=\,2) produces an irregular synthesized beam for core-dominated arrays like the SKAO. Here we adopt at most robust\,=\,1, which yields a more Gaussian PSF at comparable resolution and sensitivity.} with increasingly broad Gaussian taper at higher frequencies. However, both scale dependent sensitivity \citep[see Figs. 10/11 in][]{braun19} and tapering carry penalties for image noise. When pushing toward the low-resolution limit (e.g., in SKA-Mid Bands 5a/b) it is more advantageous in terms of final sensitivity to adjust angular resolution primarily via $uv$-weighting, and then apply as small an amount of tapering as possible. To provide an example, using the default weighting+tapering options of the sensitivity calculator: in SKA-Mid Band 1 it is possible to achieve a similar image resolution of 2.1--2.3$''$ with robust\,=\,-1 and a 1.75$''$ Gaussian taper (dimensionless tapering parameter `1' in the sensitivity calculator), or adopting robust\,=\,1 and 0.44$''$ tapering (tapering parameter value `0.25'), but the latter results in a 2.3$\times$ lower rms noise $\sigma_{\rm obs}$ in the image. When the default grid of tapering scales currently implemented in the sensitivity calculator is not fine-grained enough for comprehensive survey optimisation, we thus determine the best combination of weighting+tapering via 2D-interpolation of the available calculator output in the plane of log$_{10}\left(\sigma_{\rm obs}\right)$ vs. log$_{10}\left({\rm FWHM_{bm}}\right)$. The largest savings occur for Band 5b, where adopting robust\,=\,1 and 0.81$''$ tapering to reach our targeted angular resolution reduces observing times by nearly a factor 1.9 (due to a 40\% lower image noise) as compared to the defaults implemented in the  sensitivity calculator (robust\,=\,0; 1.89$''$ Gaussian taper). 

\begin{figure}[t!]
    \centering
	\includegraphics[width=0.52\textwidth]{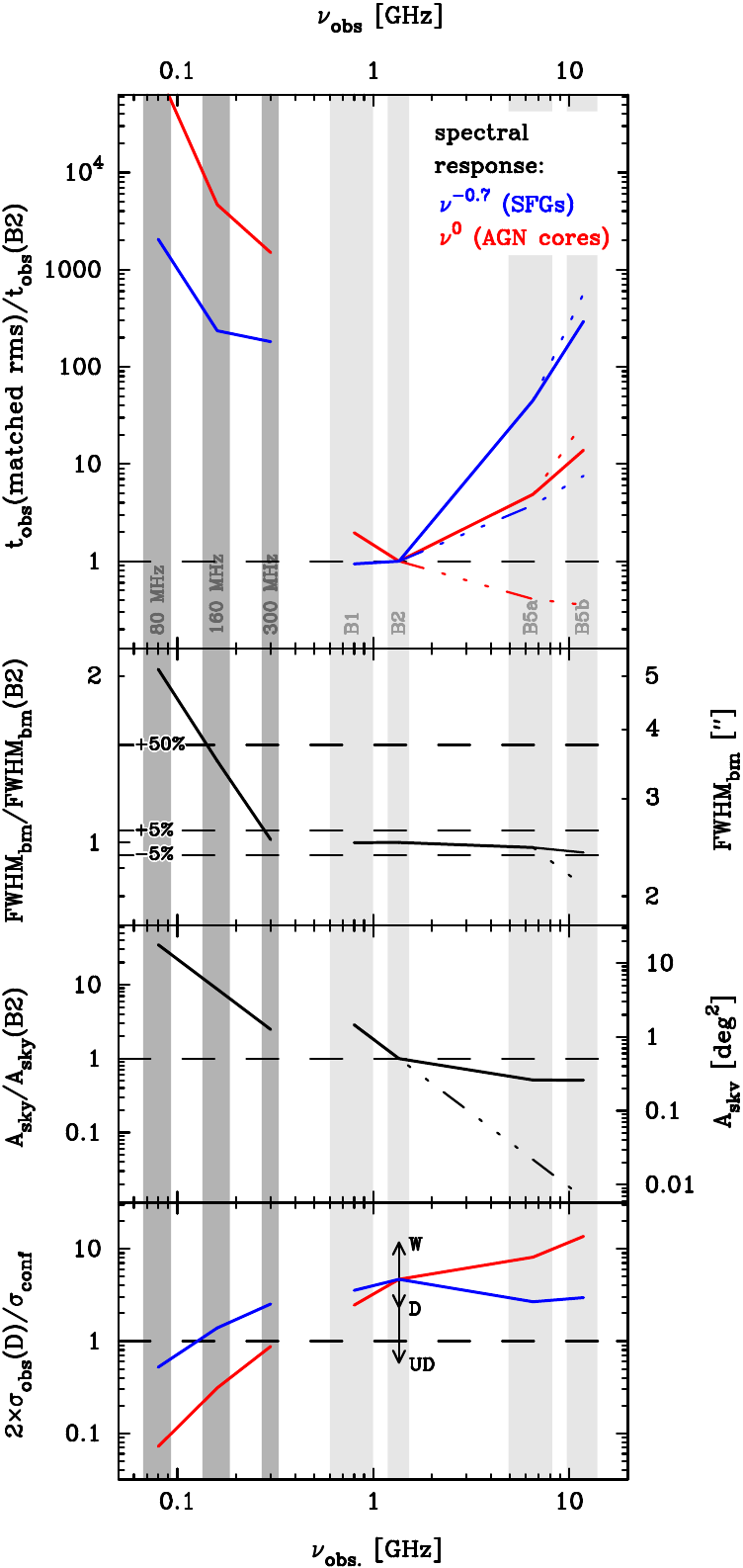}
    \caption{Panels 1--3: Band-by-band observing times ($t_{\rm obs}$), angular resolution (FWHM$_{\rm bm}$) and sky coverage (A$_{\rm sky}$) for a roughly beam-matched, multi-band AA4 survey with flat spectral response (red lines) or $\nu^{-0.7}$ scaling (blue). All three quantities are normalised to B2 (scales on left-hand y-axis). For beam sizes and sky areas, which are independent of observing time, an absolute scale is shown on the right. SKA-Low (SKA-Mid) tuning ranges with bandwidths as per Table \ref{tab:multifreqobsparams} are plotted in dark (light) grey. Solid lines in panel 1 indicate observing times for full multi-frequency coverage over an $\sim$0.25\,deg$^2$ area (cf. panel 3). Dash-dotted lines beyond B2 in panel 1 (3) denote the $t_{\rm obs}$ (A$_{\rm sky}$) for an individual pointing. The dotted line segment in row 1 (2) ends at the observing time (beam size) expected if only B5b receivers on 133 SKA-Mid dishes are used.\\
    Panel 4: band-by-band rms noise $\sigma_{\rm obs}$, for a multi-frequency survey anchored to 50\% of the sensitivity of the B2 Deep (`D') survey tier discussed in Sect. \ref{sect:B2tieredsurv}, normalised to the expected confusion noise $\sigma_{\rm conf}$. Vertical arrows illustrate the amount by which these curves would shift upward/downward for sensitivity-matched, multi-band coverage with B2 depth corresponding to the ultra-deep (`UD'), deep and wide (`W') tier.}
    \vspace{-1.5truecm}
    \label{fig:multifreq_reltsurv}
\end{figure}

Fig. \ref{fig:multifreq_reltsurv} summarizes the results of our calculations. We consider seven bands that are distributed in $\sim$0.3\,dex increments in logarithmic frequency space across the full SKA-Low and SKA-Mid tuning ranges, with the exception of a larger gap at $\sim$3\,GHz (corresponding to SKA-Mid band 3/4, both not currently part of the deployment baseline). In all bands, our aim is to cover at least 50\% of the noise-equivalent field-of-view in band 2 ($A_{eff}$, cf. Table \ref{tab:2015sur}), i.e. a minimum of 0.25\,deg$^2$. This is roughly triple the size of the combined GOODS-N/S fields, or about 10\% of COSMOS, and requires a mosaic of 12 (39) pointings in Band 5a (5b). We explore observing time requirements for two scenarios: (i) band-by-band sensitivities $\sigma_{\rm obs}$ scaling as $\nu^{-0.7}$, broadly reproducing the spectral slope of SFG radio SEDs, and (ii) a spectral response $\propto \nu^0$ akin to flat-spectrum  sources (e.g. AGN cores). Observing times $t_{\rm obs}$ for the full survey area are shown with solid lines in Fig. \ref{fig:multifreq_reltsurv} (uppermost panel), those for individual pointings at frequencies beyond Band 2 with dash-dotted lines. These observing times depend on target sensitivity and angular resolution. It is possible to closely match the angular resolution at $\sim$2.5$''$ from 300\,MHz with SKA-Low to SKA-Mid Band 5b (Fig. \ref{fig:multifreq_reltsurv}, 2nd panel). While this image resolution could be reached at 150\,MHz as well, in the `outer 24\,km' configuration and with uniform weighting, this would come at the cost of strongly increased observing times. We therefore propose to target an angular resolution of $\sim$3.5$''$ in the 160\,MHz band (see Table \ref{tab:multifreqobsparams}). This beam size remains within 50\% of the angular resolution at the higher frequencies, can be achieved in the `outer 12\,km' configuration -- which adds collecting area from an additional 66 SKA-Low stations -- and leads to an observing time commensurate with that required at 300\,MHz for a $\nu^{-0.7}$ spectral response. Note that SKA-Low has a 300\,MHz instantaneous bandwidth and can simultaneously observe at 80, 160 and 300\,MHz. Given this capability, matching observing time requirements across different SKA-Low bands represents an optimal strategy.

Total predicted survey times are dominated by the SKA-Low bands (due to sensitivity penalties when pushing the telescope to its high-resolution limit, through approximately uniform weighting and/or limitation to data from outer stations) and Band 5a/b coverage (as overcoming the intrinsically small field-of-view at high frequencies requires mosaicking).\\
Focusing first on a survey optimised for the detection of SFGs, we note that the threshold 1.5${\times}\sigma_{\rm conf}$ is reached at an rms noise of $\sim$1.9\,$\mu$Jy/bm at 160\,MHz (the band most impacted by confusion noise due to the coarser $\sim$3.5$''$ imaging\footnote{~In the following we omit from our discussions the 80\,MHz band where entering the confusion noise-dominated regime is inevitable -- for the spectral response patterns and target depths we consider here -- given the angular resolution limits of LOW.}), which for a spectral response $\propto \nu^{-0.7}$ corresponds to 0.4\,$\mu$Jy/bm in Band 2, or approximately twice shallower than the deep tier of the Band 2 survey outlined in Table~\ref{tab:2015sur}. A multi-band survey field as outlined here would most likely be located in a sky region with deep and high-quality ancillary data (e.g., also from an ultra-deep radio survey tier as the one discussed in Sect. \ref{sect:B2tieredsurv}), enabling prior based flux extraction at secure source positions in all bands, and the adoption of a lower signal-to-noise threshold at $S/N\,{\simeq}$\,3 than viable in a blind field. Despite the nominally 2$\times$ higher noise level, this multi-frequency survey would thus effectively have a very similar equivalent depth as the `Deep' tier in Table \ref{tab:2015sur}.  We report the nominal depth of the multi-frequency survey as the data point labeled `SKA-MF' in Fig. \ref{fig:SKAsurveys}. It aligns with the area--sensitivity locus of the Band 5b reference surveys in Table \ref{tab:2015sur}. Compared to current multi-frequency survey fields with spectral coverage up to 10\,GHz (X-band) or beyond -- e.g., COSMOS-XS \citep[][labelled `COSMOS-XS-X' in Fig. \ref{fig:SKAsurveys}]{algera20} or GOODS-N \citep[`GOODS-N-X']{jimenez-andrade24} -- the proposed multi-frequency survey would bring a substantial improvement in terms of both depth and area. As such it would be highly effective for characterising the radio SEDs of $M^*$ galaxies out to the end of the epoch of reionisation, detecting in its 0.25\,deg$^2$ area  $\gtrsim$100 such objects throughout the peak epoch of galaxy formation at 1\,${<}\,z\,{<}$\,3, and of order 10 $M^*$ galaxies even at $z\,{\simeq}$\,5 (see Fig. \ref{fig:SFGs_in_tiers}\footnote{~Number densities plotted in Fig. \ref{fig:SFGs_in_tiers} are for a population fraction of 20\%. For the full SFG population, the survey areas required to reach a given sample size will thus be 5$\times$ smaller than as per the lower panel of Fig. \ref{fig:SFGs_in_tiers}.}). The SEDs of LIRG-like objects (SFR\,=\,10-100\,$M_{\odot}$/yr) could be studied with hundreds of objects out to cosmic noon ($z\,{\sim}$\,2), and for MW-like objects out to $z\,{\sim}$\,1. The overall time requirement for this deep multi-band survey would amount to roughly 10$\times$ the 500-hr threshold for large project allocations (this takes into account that all SKA-Low bands can be observed simultaneously). As it facilitates studies of the SED properties in particular of those galaxies contributing most to the cosmic SFRD, a survey like this would be well suited to calibrating synchrotron radio $K$-corrections and SFRs, with a view to obtaining a precise measurement of the cosmic SFH (e.g., \citealp{An2026.SKA}), or for tracing the cosmic SFH via radio free-free emission (\citealp{Algera2026.SKA}).\\
We note that dropping the requirement of confusion-free imaging all the way down to 160\,MHz, and instead settling for just one high-resolution SKA-Low band at 300\,MHz (with 3.05$''$ resolution, via weighting with robust\,=\,-1), would reduce the overall survey time by $\sim$30\%. Due to the physical information contained in different spectral features (see, e.g., Fig. 1 in \citealp{Moldon2026.SKA}), reconstructing radio SEDs over the maximum possible frequency range would be particularly important for galaxy populations with non-standard physical conditions, e.g., extreme emission line galaxies serving as EoR analogues \citep[]{Bait2026.SKA}. However, a somewhat reduced frequency coverage might be acceptable for some scientific objectives, e.g., for projects focusing on the separation of free-free and synchrotron emission from cosmic noon galaxies.
Finally, a reduced frequency baseline limited to SKA-Mid coverage could also be sufficient for a science case targeting the spatial decomposition of free-free and thermal emission in intermediate-redshift galaxies, as described in \citet[]{Tabatabaei2026.SKA}. For this purpose, SKA-Mid data from the survey outlined above could be imaged at higher angular resolution, removing especially in Band 5a/b the tapering and accepting some scale-dependent sensitivity penalties in Band 1 to match the 1--15\,GHz coverage at a common angular resolution of $\sim$0.6$''$.

Fig. \ref{fig:AGNs_in_tiers} highlights that statistical studies of radio(-loud) AGN in massive DM halos strongly benefit from $\gtrsim$10\,deg$^2$ sky coverage. In concluding this section we thus briefly discuss the scope for obtaining resolution-matched, multi-band imaging over larger areas for AGN SED science. We do so assuming that a flat spectral response would be most suitable for such AGN-oriented multi-band coverage. Band-by-band rms noise values scaling $\propto \nu^0$ require shorter (longer) observing times at high (low) frequencies, compared to a survey with band-by-band sensitivities varying $\propto \nu^{-0.7}$. As we show in the lowermost panel of Fig. \ref{fig:multifreq_reltsurv}, a flat spectral response down to $\sim$160\,MHz can only be maintained at rms noise levels ${\gtrsim}1$\,$\mu$Jy/bm due to confusion noise in SKA-Low imaging with angular resolution 2.5-3.5$''$ (see also discussion in the previous paragraph). However, the long observing times required for deep SKA-Low coverage with this imaging resolution essentially preclude observations of more than a single pointing. On the other hand, per-pointing observing times in Bands 5a and 5b are short (0.6--0.7\,hr) when targeting an rms noise level of order 1\,$\mu$Jy at 2.5$''$. Coverage of a 10\,deg$^2$ area in Band 5a (5b) could thus be achieved with a total of $\sim$300 (900) hours, and could be combined with Band 2 and/or Band 1 coverage from wide-area imaging obtained in the context of other galaxy evolution (see, e.g., Sect.~\ref{sect:B2tieredsurv}) or cosmology-focused projects. This would provide a complementary, shallower tier with frequency coverage over 0.6--15\,GHz to accompany a deep 0.25\,deg$^2$ multi-band survey optimised for SFG studies.

\subsection{Multi-epoch observations of variability in active galaxies}
\label{sect:variability_scdrivers}

With integration times from 10s to 1000s of hours per pointing (see Table \ref{tab:2015sur}), the surveys above are well suited to implementing multi-epoch observing strategies and exploring AGN variability. Particularly interesting is the case of low-luminosity AGN populations, where there are few such studies in the radio domain. Variability not only provides valuable insights into AGN physics, it also provides an additional diagnostic to pinpoint AGN cores embedded in SFGs. In the following, we provide an overview  of recent findings and current understanding in this field.

\subsubsection{Radio Variability in Seyfert Galaxies}

Although radio monitoring of Seyfert galaxies has been limited \citep{LalShastri2011}, available multi-epoch studies indicate that radio variability is both frequent and intrinsic to their active nuclei. 
For example, \citet{Mundell2009} reported variability in roughly 50\% of a 12-source Seyfert galaxy sample observed at 8.4\,GHz over a 7-yr baseline, despite the objects not being selected for any known radio activity.
The detected fractional flux density (ratio of core emission to extended radio emission) variations reached tens of percent for at least one source; e.g., NGC 2110 showing a $>$38\% decline between epochs.
These changes in flux density are likely lower limits, as sparse two-epoch sampling preferentially detects slowly varying or fading sources; repeated flaring on $\sim$5-yr timescales, as in III Zw 2, would produce still larger peak-to-peak changes.
Often radio variability is seen to be confined to compact or unresolved cores, implying an origin in the innermost jet base or nuclear synchrotron region, whereas sources with resolved jets remained steady and showed no radio variability.
The radio cores of Seyfert galaxies that display radio variability typically have 8.4\,GHz luminosities in the range $8.2{\times}10^{18}\,{<}\,L_{\rm {core,\,8.4\,GHz}}\,{<}\,5.9{\times}10^{21}$\,W/Hz. This year-long timescale of radio variability is consistent with shocks or expanding components within compact jets. These results also suggest that radio variability is a common and characteristic property of Seyfert nuclei, reflecting episodic energy release from the central engine.

\subsubsection{Radio Variability in Low-Ionization Nuclear Emission-line Regions (LINERs)}

LINERs are the most common class of low-luminosity AGN \citep{Heckman1983,Keel1984}, found in $\sim$30\% of nearby galaxies, and are often considered a transition population between normal and Seyfert galaxies.
Their nuclear activity has been studied in the past with high-resolution VLA imaging of compact radio cores, a defining feature of AGN-powered LINERs \cite[e.g.,][]{Nagar2002,Nagar2005}.
These cores typically exhibited flat spectra and luminosities of 10$^{18}$--10$^{20}$ W/Hz, but most showed only weak ($<$10\%) long-term radio variability.
The notable exception is the LINER nucleus of M81 \cite{vanDyk1998}, where multi-frequency monitoring has revealed substantial radio variability (by factor $\sim$2) on timescales of weeks to months, along with structural evolution (e.g., in VLBI-scale jet components), confirming evidence of ongoing low-level jet activity.
Such behavior supports the view that some LINERs host scaled-down analogs of Seyfert galaxy jets, though powered by radiatively inefficient accretion flows operating at much lower Eddington ratios.
Overall, radio variability in LINERs appears rarer and milder than in Seyfert galaxies, yet its presence in cases like M81 suggests that time-variable jet production may persist even in the faintest AGN, linking the two populations through a continuum of accretion and jet power.

\section{Survey feasibility and the pathway to AA4}
\label{sect:pathAAstar}

\begin{figure}[t]
    \centering
	\includegraphics[width=0.9\columnwidth]{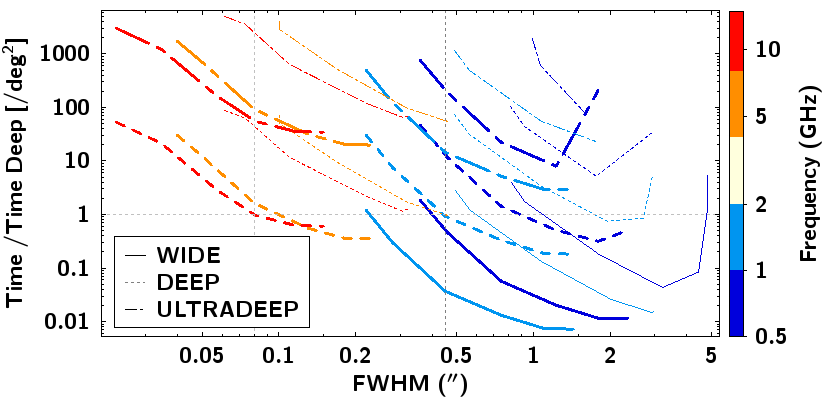}
    \caption{Integration time per sq. degree as a function of angular resolution from the SKAO Sensitivity Calculator (Decl. $=-45^{\circ}$, Elevation $=45^{\circ}$). Different colors refer to different frequencies: Band 1/0.8 GHz; Band 2/1.4 GHz; Band 5a/6.6 GHz; Band 5b/11.9 GHz. Different lines refer to different tiers: Wide (solid); Deep (dotted); Ultra-deep (dot-dashed). Thick (thin) lines refer to AA4 (AA$^*$). Band 1 and 2 (Band 5a and 5b) exposure times are normalized to the time per sq. degree reported for the Band 2 (Band 5b) deep tier in Table \ref{tab:2015sur}, i.e 304\,hrs (715\,hrs). The growing impact of confusion noise at coarser resolution results in an exponential increase of the exposure time in some cases. For Band 5b estimates we assume that also MeerKAT antennas are equipped with Band 5b receivers (see Sect. \ref{sec:multif2} for a discussion on the effects on sensitivity and angular resolution for AA4 and AA*).}
    \label{fig:time_FWHM}
\end{figure}

\begin{table}
\footnotesize
\centering
  \caption[Landscape table]{Representative Band 2 survey strategies for full (AA4) and limited (AA$^*$) telescope capabilities.}  
  \label{tab:scenarios}
 \begin{tabular}{lcc|cc|cc}
\multicolumn{7}{c}{\bf A -- AA4 capabilities}\\[1ex]
\hline
\multicolumn{3}{c|}{} & \multicolumn{2}{c|}{\bf A.1 -- Full Area} & \multicolumn{2}{c}{\bf A.2 -- 50\% Area}\Tstrut\Bstrut\\
\hline
Tier & rms & t/deg$^2$ & Area &   t$_{total}$ & Area &   t$_{total}$\Tstrut\\ 
   &   [$\mu$Jy/bm] & [hr] &  [deg$^2$] & [hr] &  [deg$^2$] & [hr]\\[1ex]
        \multicolumn{1}{c}{(1)} & \multicolumn{1}{c}{(2)} & (3) & (4) & (5) & (6) & (7)\Bstrut\\
   \hline
 \multicolumn{1}{c}{\parbox{1cm}{Ultra- Deep}} & 0.05 & 4849 & $1$ &  4849 & $0.5$ & 2424\Ttwostrut\\[2.5ex]
 \multicolumn{1}{l}{Deep} & 0.2 & 304 & 20 & 6080 & 10 & 3040\\[1.5ex]
  \multicolumn{1}{l}{Wide} & 1 & 12.2 & \multicolumn{1}{c}{1000} & 12200 & 500 & 6100\Bstrut\\
\hline
\multicolumn{7}{c}{}\\
\multicolumn{7}{c}{\bf B -- AA$^*$ capabilities}\\[1ex]
\hline
\multicolumn{3}{c|}{} & \multicolumn{2}{c|}{\bf B.1 -- Full Area} & \multicolumn{2}{c}{\bf B.2 -- 50\% Area}\Tstrut\Bstrut\\
\hline
  \multicolumn{1}{c}{\parbox{1cm}{Ultra- Deep}} & 0.1 & 1212 & $1$ &  1212 & $0.5$ & 606 \Ttwostrut\\[2.5ex]
  \multicolumn{1}{l}{Deep} & 0.4 & 76 & 20 & 1520 & 10 & 760\\[1.5ex]
   \multicolumn{1}{l}{Wide} & 2 & 3.05 & \multicolumn{1}{c}{1000} & 3050 & 500 & 1525\Bstrut\\
\hline\\[-2ex]
  \end{tabular}
\begin{minipage}{\textwidth}
    \vspace{0.2cm}
    \small
    {\it Notes:} A.1 -- {\bf  AA4, Full Area} -- this is the strategy described in Table~\ref{tab:2015sur}, which provides the figures of merit illustrated in Figs. \ref{fig:AGNs_in_tiers} and \ref{fig:SFGs_in_tiers}. A.2 -- {\bf AA4, 50\% Area} -- same as A.1, but with the  population statistics requirement relaxed from 100 to 50 sources per bin (see lower panels of Figs. \ref{fig:AGNs_in_tiers} and \ref{fig:SFGs_in_tiers}), limiting the available statistics.
    B.1 -- {\bf AA$^*$, Full Area} -- in this case we relax the sensitivity requirement by a factor of two, limiting the  survey capability to trace the lowest-luminosity and/or highest redshift populations. This strategy may be better suited for early observations with $AA^*$. B.2 -- {\bf AA$^*$, 50\% Area} -- same as B.1, but with requirement on population statistics relaxed as for A.2. In this scenario both science and statistics are more limited in scope.\newline
    Time estimates in cols. (3), (5) \& (7) assume Briggs weighting with robust\,=\,-1 and are calculated with the SKAO Sensitivity Calculator, adopting as in Table~\ref{tab:2015sur}, Decl. $=-45^{\circ}$, Elevation $=45^{\circ}$, a reference frequency of 1.35\,GHz and a bandwidth of 0.34\,GHz. 
    \end{minipage}
\end{table}

Fig. \ref{fig:time_FWHM} shows the integration time needed to cover 1\,deg$^2$ at uniform sensitivity at the depth of the tiers defined in Table~\ref{tab:2015sur} for AA4 (thick lines) and AA$^*$ (thin lines), as a function of the target angular resolution. Different colors represent different bands. Band 1/2 and Band 5a/5b exposure times are normalized to the values reported in Table~\ref{tab:2015sur} for the Band 2 and 5b deep tiers (with angular resolutions FWHM\,=\,0.45$''$ and 0.08$''$, respectively). It is clear that Band 2 (Band 5b) provides a better performance than Band 1 (Band 5a), particularly at finer resolutions. Exposure times become more similar going to coarser resolutions, where higher-frequency bands are less effective due to i) their smaller fields of view (resulting in a larger number of pointings per sq. degree), and ii) stronger down-weighting of long baselines. At resolutions of ${\gtrsim}0.1''$, Band 5a and 5b performance is similar, but only for AA4. For AA$^*$ Band 5b remains advantageous with respect to Band 5a.\\
For the reasons explained in Sects. \ref{sect:B2tieredsurv} and \ref{sec:b5res}, sub-arcsec resolutions are desirable for galaxy/AGN co-evolution studies, but exposure times increase at finer angular resolution. As shown in Fig. \ref{fig:time_FWHM}, resolutions of ${\sim}0.5''$ in Band 2 and ${\lesssim}0.08''$ in Band 5b provide a good compromise between scientific and observing time requirements for AA4. Nevertheless, the total exposure times required to carry out  the designed multi-tier survey with AA4 are demanding: typically between $(5{-}10){\times}10^3$ hours per tier in Band 2, and ${\sim}3{\times}10^3$ hours per tier in Band 5. As shown in Figs.~\ref{fig:AGNs_in_tiers}~and~\ref{fig:SFGs_in_tiers}, some fine tuning in the area covered by Band 2 tiers is possible at the expense of source population statistics, while maintaining the designed flux limit requirements is necessary to probe the full range of source parameters as desired.  

During the initial phase of SKAO operations in AA$^*$, SKA-Mid will achieve a factor ${\sim}4$ poorer image resolution compared to AA4, implying larger contributions of confusion noise. This will limit the achievable sensitivity, rendering the Band 2 ultra-deep tier unfeasible. AA$^*$ could thus be used to conduct wide/deep tiers in Band 2 and/or deep/ultra-deep tiers in Band 5 at a factor 3--4 coarser resolution (to maintain similar exposure times to AA4). However, as illustrated in Fig.~\ref{fig:EDFNmultires}, it is not advisable to relax beam size much beyond FWHM\,${\sim}\,2''$, as the contribution from confusion noise becomes important at approximately the depth of the deep tier at this angular resolution (see also Fig. \ref{fig:multifreq_reltsurv} and discussion in Sect. \ref{sec:multif2}). One possibility could thus be to start from the wide tier that -- as shallower -- is less affected by source blending. Additionally, one could exploit Band 5 to reach deep and ultra-deep sensitivities at sub-arcsec (${\sim}0.5''$) resolution, and start exploring the sub-$\mu$Jy radio sky over limited survey areas (${\lesssim}1$\,deg$^2$). Alternatively, a staged approach could be adopted, in which pilot surveys are conducted at reduced sensitivity and later followed up with AA4 to reach the originally designed depth.  For illustration, Table~\ref{tab:scenarios}  presents four representative band 2 survey strategies, outlining possible scenarios under constraints in observing time and/or telescope performance (speficially, AA4 vs. AA$^*$). The area--sensitivity parameter space probed by these scenarios is also highlighted in Fig.~\ref{fig:SKAsurveys} (see pale violet, diagonal lines).

Multi-frequency survey projects aiming for a $\sim$2.5$''$ image resolution (see Sect. \ref{sec:multif2}) to reconstruct galaxy-integrated MHz-to-GHz SEDs would probe a noise regime commensurate with the Deep rather than the Ultra-deep tier, and would thus — at least in parts — be well suited to AA$^*$ capabilities. We could envisage the following scenario for distributing the multi-band imaging across both AA$^*$ and AA4. As illustrated in Fig. \ref{fig:multifreq_reltsurv}, Band 1/2 observing times constitute a small fraction of the total time requirement for a resolution- and (roughly) area-matched multi-frequency survey. Even though AA$^*$ integration times are of order twice as long as in AA4 at fixed depth, anticipating the Band 1 and 2 coverage would thus only represent a minor increase of the overall time required for a multi-frequency project. It would also constitute close-to-optimal usage of AA$^*$ capabilities, in that survey speed is maximized for angular scales 1–2$''$ in AA$^*$ (see Fig. \ref{fig:time_FWHM}). There is a strong rationale for additionally obtaining at least partial coverage at high frequencies (Band 5) of such a multi-frequency field during AA$^*$ . This will permit SED studies over $\sim$1\,dex already in the early phases of SKAO, with a focus on disentangling and cross-calibrating thermal and non-thermal emission for cosmic SFH studies \citep[e.g.,][]{Algera2026.SKA,An2026.SKA}. Once we take into account the availability of INAF/MPG Band 5b receivers also on MeerKAT dishes, it is clear that high-frequency coverage in AA$^*$ in relative terms is cheaper to obtain in Band 5b than Band 5a. For the latter, AA$^*$ observing times are approximately 3-fold longer than in AA4, while for Band 5b the increase is only a factor ${\sim}$2 according to the SKAO sensitivity calculator\footnote{~In Sect. \ref{sec:multif2} we provided a revised estimate for AA4 integration times in Band 5b, once additional Band 5b feeds on MeerKAT dishes are taken into account (factor 0.52 reduction). With MeerKAT dishes contributing a larger fraction of the total collecting area, the performance increase relative to the default sensitivity calculator output is even more significant in the AA$^*$ configuration at a factor 1.65 (area) and 0.37 (observing time).}. Moreover, this factor 2 might in practice be reduced somewhat more, as — due to the shorter $B_{\rm max}$ -- less tapering is required in AA$^*$ than in AA4 to obtain a $\sim$2.5$''$ image resolution.\\
Finally, we note that the SKA-Low element of a resolution-matched multi-band survey at the $\sim$2.5$''$ scale would necessarily have to be implemented in AA4. Pushing the resolution of the SKA-Low telescope in this way by restriction to, or aggressive weighting toward, outer stations (see Sect. \ref{sec:multif2}) implies that maximizing collecting area is indispensable. Specifically, the `outer 12\,km' configuration in AA4 offers nearly twice as many stations as AA$^*$ (102 vs. 54).

\section{Synergies}
\label{sec:synergies}

The transformational potential of SKA extragalactic continuum surveys, particularly for galaxy/AGN co-evolution studies discussed in this chapter, will only be fully realised through synergies with other major facilities. 
A full review of all possible synergies is beyond our scope. In this section we thus focus on the role of ongoing and future optical/NIR surveys, and we discuss the role of the ALMA Wideband Sensitivity Upgrade (WSU).

\subsection{Optical/NIR surveys}
\label{sec:synergies-optNIR}

As highlighted in the Introduction, a panchromatic approach is essential for a comprehensive understanding of the complex physical processes regulating galaxy growth. 
SKAO surveys will need to be complemented by observations at shorter wavelengths to fully unlock their scientific potential. Multi-band photometry and optical/near-infrared (NIR) spectroscopy will play a key role in this respect. Both are essential for: i) properly identifying radio sources and deriving source distances (through spectroscopic or photometric redshifts); ii) classifying radio sources into AGN and SFGs; and iii) inferring important physical characteristics (for example, bolometric luminosities, stellar masses, SFRs, metallicities, environment, etc.), and linking these to the radio-derived galaxy properties, such as radio SFR and AGN radio power.

We identify two primary timescales of SKAO operations:

\begin{itemize}
    \item \underline{The AA$^*$ timescale (approximately 2025-2035):} During this initial phase, SKA surveys will operate alongside powerful new survey facilities like ESA's Euclid mission\footnote{https://www.euclid-ec.org/}, the Vera Rubin Observatory\footnote{https://rubinobservatory.org}, and NASA’s Nancy Grace Roman Space Telescope\footnote{https://roman.ipac.caltech.edu} (Roman), as well as the James Webb Space Telescope\footnote{https://science.nasa.gov/mission/webb} (JWST) and ESO advanced optical/IR multi-object spectrographs (MOS; e.g., 4MOST\footnote{https://www.4most.eu/cms/home/}, MOONS\footnote{https://vltmoons.org/}). 
    \item \underline{The AA4 timescale (approximately >2035):} In its final design configuration, SKAO will synergise with the next generation of flagship facilities, including the Extremely Large Telescope\footnote{https://elt.eso.org/} (ELT), and concepts like the Wide-Field Spectroscopic Telescope\footnote{https://www.wstelescope.com/} (WST).
\end{itemize}

In the AA$^*$ period multi-wavelength data will provide essential information for distinguishing between physical processes (e.g., SF vs. AGN) in sources that the SKAO may not yet spatially resolve. A comprehensive overview of the  synergies between the SKAO and ESO MOS instruments for galaxy/AGN co-evolution studies is presented in \citet{Prandoni2024}. Here we  briefly discuss the unique role of the new generation of optical/NIR space missions, like Euclid (launched in 2022) and the soon-to-be-launched Roman Space Telescope\footnote{roman.ipac.caltech.edu}.

Euclid surveys will dramatically change the astronomy landscape in the 2030s. While designed for cosmology, they will also be invaluable for assessing the role of
environment on galaxy formation and evolution. Euclid spectroscopy will provide an unprecedented view of the large scale structure (LSS) at cosmic noon, enabling clustering analyses of different galaxy/AGN populations. Euclid high resolution (0.2--0.4$''$) optical and near-infrared photometry will provide direct estimates of the DM halo mass and distribution around galaxies through weak lensing. The combination of Euclid imaging with multi-band, wide-field optical photometry from the Legacy Survey of Space and Time (LSST) at the Vera Rubin Observatory will be particularly impactful; more precise photo-zs benefit galaxy evolution studies, and smaller photo-z errors increase the weak lensing signal (\citealt{Rhodes2017}). Euclid will therefore play a unique role for investigating the SHMR at redshift $z\,{>}$\,1, and, combined with SKA surveys,  how it is influenced by jet-driven AGN feedback (see Sect.~\ref{subsec:shmr}). 

Euclid flagship surveys will cover most of the extragalactic sky (14,700\,deg$^2$; wide survey), plus 53\,deg$^2$ over three so-called Euclid Deep Fields (EDFs), which will be visited multiple  ($>40$) times over 6 years, providing spectroscopic redshifts down to e.g. F(H$_{\alpha)}\,{\sim}\,5{\times}10^{-17}$\,erg\,s$^{-1}$\,cm$^{-2}$ (i.e. SFR\,${\sim}3M_{\odot}$/yr at $z\,{\sim}$\,1), and robust photometric redshifts from multi-band (IYJH + ugrizy from LSST) imaging to $H_{\rm AB}$\,=\,26.4 (5$\sigma$ flux limit), i.e. 2\,mag deeper than the wide survey. Two EDFs (EDF-Fornax covering 10\,deg$^2$ and the 23\,deg$^2$ EDF-South) are accessible to the SKA telescopes and would therefore be ideal targets for the deep tier of the surveys outlined in Table~\ref{tab:2015sur}.

Roman will significantly extend Euclid’s capabilities thanks to its higher sensitivity, faster survey speed, and superior ($0.1''$) resolution. Roman will conduct multiple surveys using its Wide Field Instrument.  
The recently finalized Roman Core Community Survey program (\citealt{Zasowski2025}) includes two major extragalactic components:

The {\it High Latitude Wide Area Survey}
\begin{itemize}
\item Medium/wide tiers: Multi-filter imaging and grism spectroscopy over ${\sim}$2,400\,deg$^2$ (medium), reaching $\sim 26.5\,$AB-mag ($5\sigma$) and $1.5{\times}10^{-16}$\,erg s$^{-1}$\,cm$^{-2}$ ($5\sigma$). Another ${\sim}$2,700\,deg$^2$ will be imaged in a single band to $H_{AB}=26.2$ (wide). These tiers overlap with Euclid wide surveys. The medium tier includes Euclid’s southern deep fields.
\item Deep/ultra-deep tiers: Multi-band imaging and spectroscopy of the COSMOS and XMM-LSS Rubin deep-drilling fields, reaching 1\,mag deeper than the medium tier (deep). The ultra-deep component (5\,deg$^2$ imaged in YJH) will go another 0.5\,mag deeper. 
\end{itemize}

The {\it High Latitude Time Domain Survey (ELAIS-N1 and EDF-South) } 
\begin{itemize}
\item Wide/deep imaging tiers: 10.68\,deg$^2$ (North) and 7.59\,deg$^2$ (South) in RzYJH with ${\sim}$10-day cadence (wide); 1.97\,deg$^2$ (North) and 4.5\,deg$^2$ (South) in zYJHF with interlaced sequences (deep), probing objects up to ${\sim}$10 billion years old.
\item Wide/deep spectroscopy tiers: 4.5\,deg$^2$ and 0.56\,deg$^2$ within the southern deep imaging region (EDF-South), with ${\sim}$5-day cadence.
\end{itemize}

As for Euclid, Roman surveys will be excellent targets for the SKA wide, deep and ultra-deep tiers.

\begin{figure}[t]
    \centering
	\includegraphics[width=0.9\columnwidth]{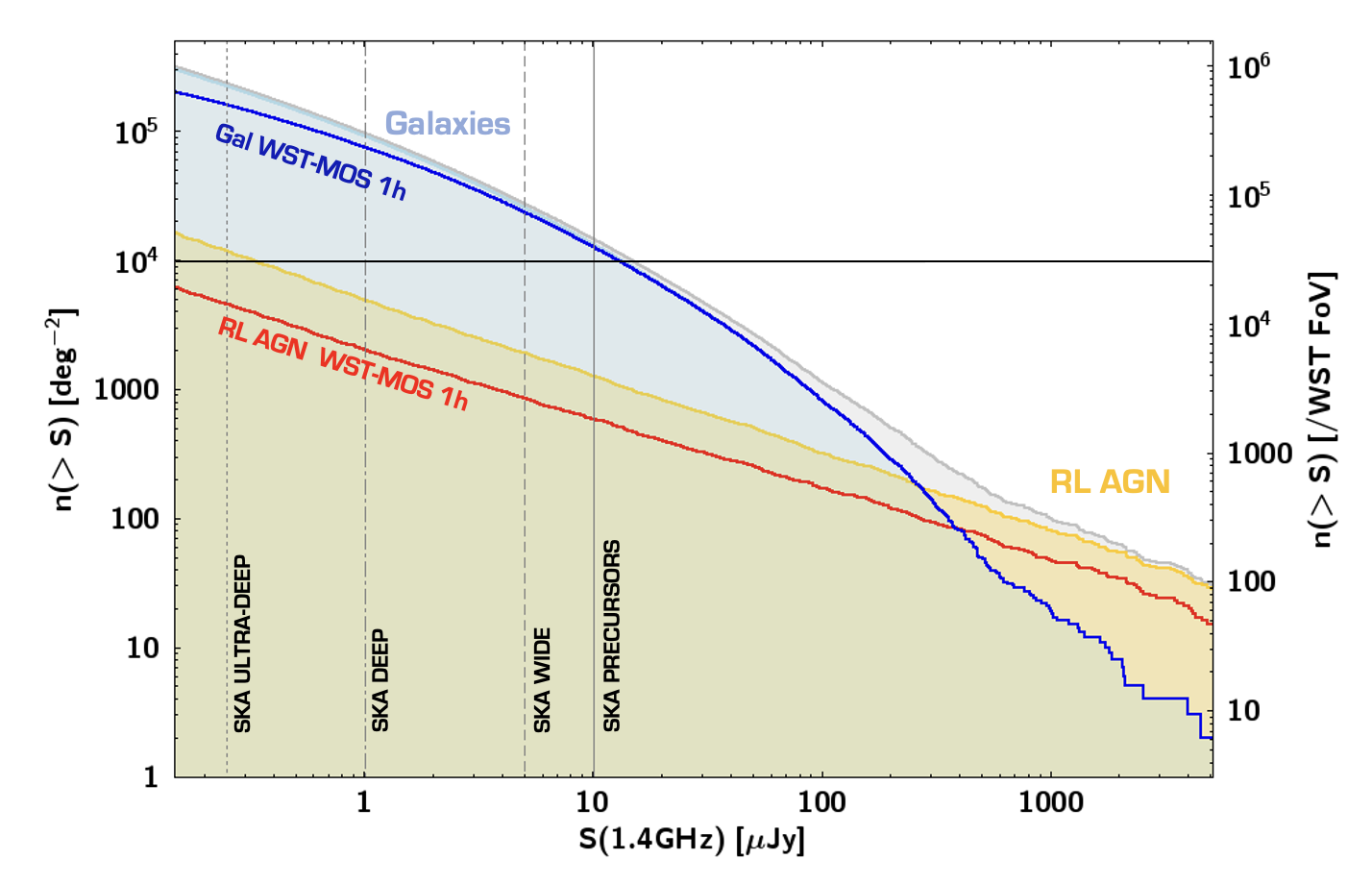}
    \caption{Cumulative number density predictions for different classes of radio continuum sources from T-RECS (\citealt{bonaldi23}), compared with WST capabilities. left-hand y-axis -- number of sources in a 1\,deg$^2$ field; right-hand y-axis -- number of sources in the WST field of view. The black horizontal lines indicate the number of fibres available for WST (30k for low resolution spectroscopy). The filled histograms represent radio source populations: galaxies (cyan), radio AGN (orange) and the sum of the two (light grey). The thick blue/red lines correspond to sources from the aforementioned populations that can be detected by WST-MOS in a 1-hr exposure (\citealt{Bacon2024}). The vertical lines are 5$sigma$ radio continuum depths for SKA-Mid band 2 survey tiers (Table~\ref{tab:2015sur}) and MeerKAT/MIGHTEE. Adapted from \citet{Prandoni2024}.}
    \label{fig:WST_SKA}
\end{figure}

Over the next decade, imaging from a large number of survey-oriented ground-based (e.g., Rubin/LSST, SKAO, CTA, Einstein Telescope) and space-based (e.g., Euclid, Roman, LISA) facilities will detect an enormous number of celestial objects with unprecedented precision. To fully characterise and understand these, spectroscopic follow-up at adequate spectral resolution and cadence is required. Given the expected number of sources, a dedicated wide-field spectroscopic facility is needed to capitalise on the huge investment made in these imaging facilities. The Wide-field Spectroscopy Telescope (WST) concept (\citealt{Bacon2024}), that will be proposed as the next ESO project after completion of the ELT, could fulfill this role.  WST is a 12-metre wide-field spectroscopic survey telescope with simultaneous operation of a large field-of-view (3\,deg$^2$), highly multiplexed (30,000) MOS, with both low- and high-resolution modes, and a giant 3.3\,arcmin$^2$ integral field unit. As shown in Fig.~\ref{fig:WST_SKA}, in 1 hour of integration time, WST-MOS will be able to detect and provide spectroscopy for virtually all galaxies detected down to the ultra-deep tier of the SKA surveys outlined in Table~\ref{tab:2015sur}, as well as for about 50\% of the radio loud AGN populations.\\
Combining SKAO AA4 observations with WST surveys will offer a uniquely powerful, multi-dimensional view of galaxy evolution and LSS. SKAO’s sensitive radio data will probe SF, gas content, and AGN activity, while WST will provide precise redshifts, kinematics, and chemical information for millions of galaxies. Together, they would enable accurate cross-matching, improved environmental studies, and transformative constraints on cosmology by linking baryonic and dark-matter tracers across cosmic time. This synergy could significantly enhance the scientific return of both facilities and open new discovery space that is inaccessible to either alone.

Though not designed to be survey instruments, facilities like JWST and in the future ELT can provide important complementary information for SKAO-based galaxy/AGN co-evolution studies. With SKAO playing a primary role in detecting radio AGN in the EoR (see \citealt{Afonso2026.SKA}, for a more comprehensive discussion), JWST adds deep, high-resolution infrared imaging and spectroscopy to reveal the earliest galaxies ($z\,{\sim}$\,10 and beyond), and ELT will provide unmatched spatial resolution and sensitivity from the ground for detailed kinematics, metallicity, and resolved structures of high-$z$ populations. Together with the SKAO, these facilities will allow us to connect gas, dust, stars, and SMBH growth within the same galaxies across cosmic time, enabling precise measurements of SF histories, feedback processes, and early galaxy assembly that none of the observatories could fully achieve on their own. 

\subsection{Synergies between ALMA WSU and SKAO}

Combined observations with the SKAO and Atacama Large Millimeter/submillimeter Array Wideband Sensitivity Upgrade (ALMA WSU) will usher in a new era of extragalactic radio and (sub-)~millimetre astronomy, providing a panchromatic and physically comprehensive view of galaxy-SMBH co-evolution. 
This synergy is defined by the complementary roles of the two facilities: SKAO as the unbiased survey machine offering statistical power across vast cosmic volumes, and ALMA WSU as the high-resolution follow-up
instrument, essential for dissecting the molecular gas content and kinematics of selected sources \citep{Carpenter2022,ALMAWSUFactsheet}.

The goal is to implement a ``survey-to-pointed'' strategy, using SKAO’s dust-unbiased census of SFGs and AGNs to select statistically robust samples for detailed molecular analyses with ALMA. This approach maximizes the scientific return by linking large-scale population studies (SF history, AGN demographics) to small-scale physical processes (gas flows, feedback mechanisms; \citealt{Prandoni2015,coogan18}). The integration of these capabilities is particularly effective because the ALMA WSU deployment -- with initial scientific observations targeted toward the end of the decade 
and full capabilities expected in the early 2030s \citep{Carpenter2022,ALMAWSUpage} -- aligns well with the SKAO staged delivery plan. 

The ALMA WSU represents a fundamental transformation, enhancing ALMA’s instantaneous bandwidth and observing efficiency by upgrading receivers, digitizers/digital transmission and the correlator \citep{Carpenter2022,ALMAWSUFactsheet}. These improvements are crucial for SKAO synergy because they dramatically reduce the observation time required for deep, targeted studies that trace the cold gas reservoir in high-z galaxies selected by SKAO.
The ALMA WSU will deliver substantial gains by increasing continuum mapping speed by a factor of $\sim$4.8 or more, and spectral scan speeds by a factor of 10 or more\footnote{~The WSU substantially reduces observing time, but does not remove the need for careful planning of correlator resources, data rates, and long-term archiving. Observatory computing and pipeline upgrades remain a necessary component of achieving the advertised science performance \citep{ALMAWSUFactsheet}.}. Crucially, the upgrade will also largely eliminate the traditional trade-off between bandwidth and spectral resolution, vastly improving the efficiency of spectral surveys (see \citealt{Carpenter2022,ALMAWSUFactsheet}), and transforming ALMA in a very efficient "redshift machine". Indeed ALMA-WSU will observe spectral lines from (obscured) distant galaxies with greater speed, allowing astronomers to confirm their existence and study their properties, more quickly and with greater detail than before.

In the following we present two general science cases that would greatly benefit from a synergistic use of the SKAO and the ALMA WSU.

\subsubsection{Star Formation and Molecular Gas Reservoirs}

A central scientific goal is to obtain a complete, dust-unbiased measurement of the Cosmic Star Formation History (SFH) and to understand how the relationship between SFR and molecular gas mass ($M_{\rm mol}$) evolves across cosmic time.\\
While SKA continuum observations provide a unique, dust-unbiased probe of SF, converting radio luminosity into an SFR nevertheless requires care: multi-frequency observations (see Sect. \ref{sec:multif}) are indispensable for separating thermal and non-thermal SED contributions and for robust SFR estimates \citep[e.g.,][and \citealp{Algera2026.SKA}]{Murphy2011,taba_17}.

This component separation is potentially further complicated by a third process. \textit{Planck} measurements of the Milky Way indicate that Anomalous Microwave Emission (AME) can be a significant contributor at 10s of GHz in SF environments \citep{Dickinson2018}. Although AME is well-studied in our Galaxy, it is largely unexplored in extragalactic sources, for which SKAO will open a new frontier of AME studies. This creates a powerful synergy with ALMA, as accurately reconstructing the full radio SED to disentangle synchrotron, free-free, and AME emission requires the high-frequency constraints provided by ALMA observations \citep[see, e.g., Fig. 6 of ][]{Dickinson2018}. For galaxies at $z\,{\gtrsim}$\,1 neglecting this component could therefore bias the interpretation of Band 5 observations, which are critical for tracing the thermal emission.\\
The ALMA WSU provides the detailed gas kinematics and mass measurements (via multiple CO transitions, dust continuum, and atomic fine-structure lines) needed to infer SFE and to map gas flows and feedback in AGN host galaxies \citep{coogan23,ALMAWSUScience}.

\paragraph{SKAO Survey Selection}
This case primarily utilizes the Band 2 deep and ultra-deep tiers, as well as the Band 5b ultra-deep tier outlined in Table \ref{tab:2015sur}. The former can be used to trace SFGs up to high redshifts (see Fig. \ref{fig:SFGs_in_tiers}), the latter provides the ability to map SF distributions on sub-kpc scales to intermediate redshifts 
\citep[see also][]{Murphy15} and to probe regimes where free-free emission contributes significantly to the radio SED, enabling more direct SFR estimates. 

\paragraph{Staged science delivery}
During the AA$^*$ phase, SKA deep surveys will identify large samples of SFGs at cosmic noon ($z\,{\sim}$\,1--2). ALMA WSU (with initial 2-fold BW increase) will follow up these samples efficiently. The increased continuum speed (representative factor $\sim$4--5 in some bands for BW$\times$2) enables rapid acquisition of dust continuum measurements and simultaneous detection of multiple CO transitions for robust $M_{\rm mol}$ estimates for statistical samples \citep{Carpenter2022,ALMAWSUFactsheet}.\\
With SKA-Mid AA4 design baseline capabilities, ultra-deep tiers will extend the radio census to earlier and fainter galaxies (as illustrated by, e.g., Fig. 2 in \citealp{coogan23}, detection limits in terms of SFR depend on assumptions about the radio--SFR conversion; under standard assumptions populations with SFRs of order 10-100\,$M_{\odot}$/yr will be detectable up to $z\sim6$ -- see Fig.~\ref{fig:SFGs_in_tiers}). ALMA WSU in its final phase (BW$\times$4) will be used to identify and map FIR fine-structure lines such as [C\,II] and [O\,III] in the early Universe, although the efficiency of blind fine-structure-line surveys depends on line luminosity functions. Joint SKAO--ALMA datasets will robustly test the radio--FIR correlation at high redshift provided that AGN contamination and cosmic-ray physics are properly accounted for \citep{Carpenter2022,Braun2015}.

\begin{figure}[t]
    \centering
    \includegraphics[width=0.8\textwidth]{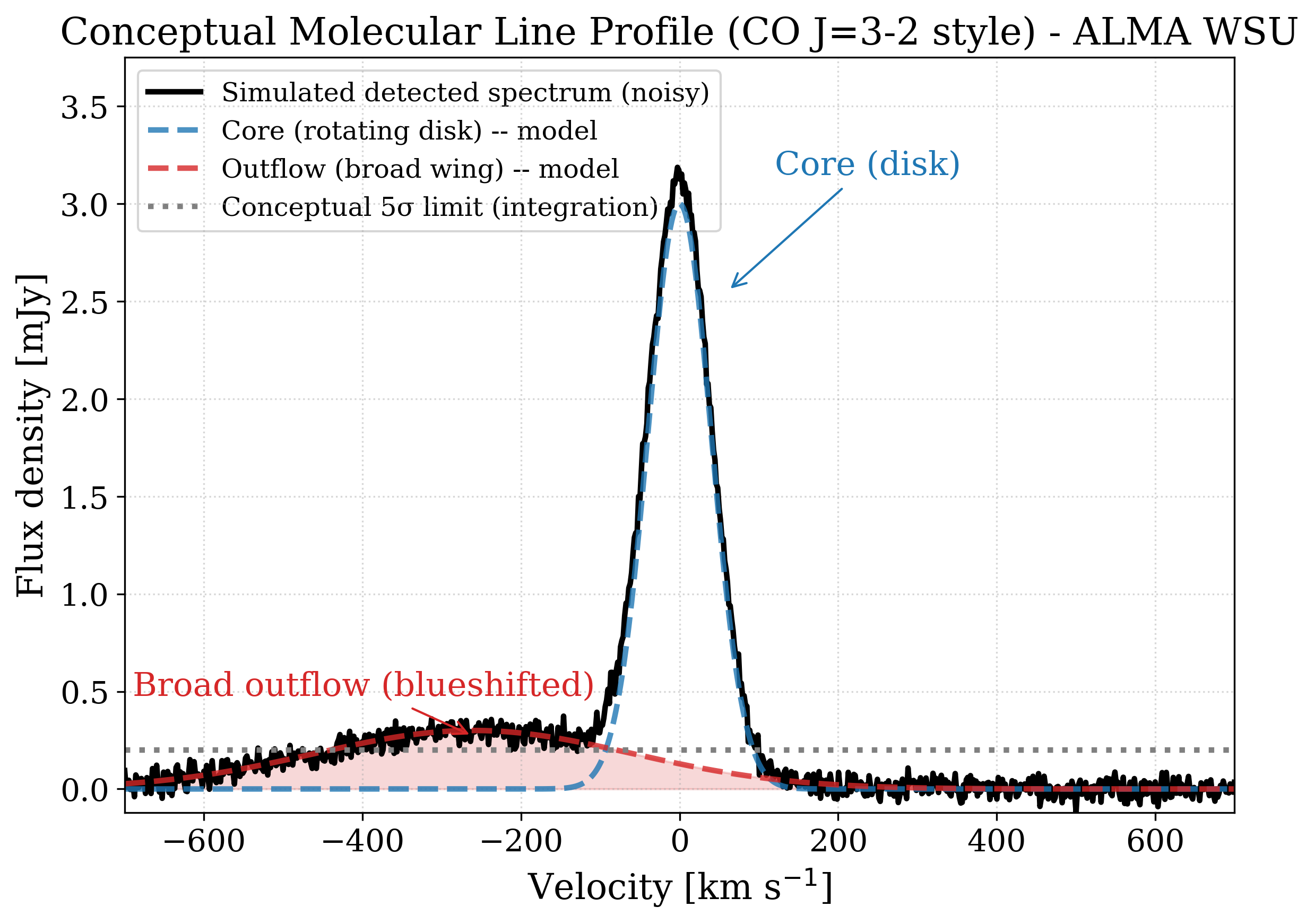}
\caption{Simulated ALMA WSU spectral profile of a molecular transition (e.g.\ CO $J$=3--2). The model consists of a narrow core component (rotating disk; Gaussian FWHM $\sim$90\,km\,s$^{-1}$; peak $\sim$3\,mJy) plus a broad, blueshifted outflow wing (peak $\sim$0.3\,mJy; width $\sim$200\,km\,s$^{-1}$). 
The black trace shows the noisy realization (added Gaussian RMS = 0.04\,mJy); the dotted horizontal line marks a conceptual $5\sigma$ threshold for the chosen integration. 
\emph{Caveats:} fluxes and noise are illustrative; in practice detection of faint outflow wings depends on integration time, spectral resolution, line luminosity, stacking strategy and accurate baseline subtraction.}
\label{fig:COoutflow}
\end{figure}

\subsubsection{Dissecting AGN Feedback and Galaxy--Black Hole Co-Evolution}

Understanding how AGN fueling/feedback cycles regulate SF and galaxy growth is key to galaxy evolution theory. Important insights in this regard may come from spatially-resolved, multi-component (stars, neutral, molecular and ionized gas) investigations of the gas kinematics and distribution in jetted AGN on all scales (from the innermost sub-kpc regions to the surrounding inter-galactic medium). The SKAO can provide a complete census of jetted AGN across luminosity and morphology, as well as information on the distribution and kinematics of the neutral hydrogen component, that traces the bulk of the cold gas in galaxies, and is the most abundant gas phase in the inter-galactic medium. ALMA, on the other hand, probes denser molecular gas reservoirs, which are more directly connected to star forming regions. Combining SKAO and ALMA (as well as optical integral field spectroscopy) one can fully trace multi-phase gas in- and out-flows in jetted AGN, across a range of physical scales, as well as gather information on the gas physical conditions, and ultimately on the impact  of jets in regulating star formation in the AGN host galaxy (for more details see \citealt{Maccagni2026.SKA}). Radio continuum observations are sensitive to jetted AGN independently of optical/IR obscuration; however, morphological and spectral diagnostics are needed to separate AGN-related radio emission from SF, particularly at galactic and sub-galactic scales (see Sects.~\ref{sec:angres}, \ref{sec:multif} and \ref{sec:synergies-optNIR}). 

\paragraph{SKAO Survey Selection}
This case benefits from both the SKA-Mid tiered, multi-resolution surveys presented in Table~\ref{tab:2015sur} and the multi-frequency strategy of Sect.~\ref{sec:multif2}. In particular, deep multi-frequency analysis enables the measurement of the radio spectral index, allowing us to distinguish between optically thin (steep) and optically thick (flat/inverted) synchrotron emission, as well as between non-thermal and thermal emission. These diagnostics, combined with morphology and multi-wavelength data, help separate RL AGN, RQ AGN, and SFG contributions \citep{Condon1992}. On the other hand, $\sim$0.1$''$ resolution images are essential to distinguish compact AGN cores and jets from extended SF-related emission, enabling cleaner pre-selection for ALMA kinematic follow-ups that can reveal AGN-related outflows.

\paragraph{Staged science delivery}
SKAO AA$^*$ will provide a deep, statistical classification of $z\,{\lesssim}$\,1 RQ AGN populations (Fig. \ref{fig:AGNs_in_tiers}). ALMA WSU (BW$\times$2) will follow up selected RQ and RL AGN to map molecular gas kinematics (e.g. CO $J=3\rightarrow2$) and search for signatures of inflows and outflows (see Fig.~\ref{fig:COoutflow}). Enhanced spectral sensitivity enables detection of broad, low-surface-brightness wings associated with outflows, but careful stacking and deep integrations (and ACA/TP when necessary) are often required to robustly quantify the faintest components \citep{coogan23}.\\
With SKA-Mid AA4, ultra-deep surveys will reach the highest morphological fidelity for source pre-selection \citep{braun19,Prandoni2015}. ALMA WSU (BW$\times$4) provides the sensitivity and bandwidth to map fainter and more extended molecular outflows, helping to establish spatial correlations between central radio jets and disturbed molecular gas kinematics \citep{ALMAWSUFactsheet}.

\section{Summary}
This chapter highlights how SKA extragalactic radio continuum surveys will open a transformative window into galaxy and AGN co-evolution by delivering deep, dust-unbiased measurements across large cosmic volumes. Through multi-tier, multi-frequency and multi-resolution strategies, the SKAO will enable robust studies of the SMBH–galaxy–halo connection, probe AGN feedback across environments and epochs, and provide statistical samples capable of overcoming cosmic variance, thereby capturing a wide range of source populations, including rare classes of objects and extreme environments. This chapter also shows that multi-frequency imaging, multi-epoch observations, and synergies with major optical/NIR survey facilities -- Euclid, Rubin, Roman, and future spectroscopic observatories -- are essential for unlocking the full scientific potential of the SKAO. Finally, we demonstrated the role that ALMA-WSU can play in supporting SKA surveys. Together, these efforts will deliver an unprecedented, panchromatic view of the physical processes shaping galaxies and AGN up to cosmic noon and beyond.

\bibliographystyle{abbrvnat-maxbibnames4}

\bibliography{chapter}

\end{document}

%% file: journal-names.tex
\newcommand{\actaa}{Acta Astron.} 
\newcommand{\araa}{ARA\&A} 
\newcommand{\aar}{A\&ARv} 
\newcommand{\aapr}{A\&ARv} 
\newcommand{\ab}{Astrobiol.} 
\newcommand{\aj}{AJ} 
\newcommand{\apj}{ApJ} 
\newcommand{\apjl}{ApJL} 
\newcommand{\apjs}{ApJSS} 
\newcommand{\ao}{Appl. Opt.} 
\newcommand{\apss}{Astro. \& Space Sci.} 
\newcommand{\aap}{A\&A} 
\newcommand{\aaps}{A\&AS.} 
\newcommand{\baas}{Bull. Am. Astron. Soc.} 
\newcommand{\caa}{Chinese A\&A} 
\newcommand{\cjaa}{Chinese J. A\&A} 
\newcommand{\cqg}{Class. Quantum Gravity} 
\newcommand{\gal}{Galaxies} 
\newcommand{\gca}{Geo. Cosmo. Acta} 
\newcommand{\icarus}{Icarus} 
\newcommand{\jcap}{JCAP} 
\newcommand{\jgr}{J. Geophys. Res.} 
\newcommand{\jgrp}{J. Geophys. Res. Planets} 
\newcommand{\jqsrt}{J. Quant. Spectrosc. Radiat. Transf.} 
\newcommand{\memsai}{Mem. SAIt} 
\newcommand{\mnras}{MNRAS} 
\newcommand{\nat}{Nature} 
\newcommand{\nastro}{Nat. Astron.} 
\newcommand{\ncomms}{Nat. Commun.} 
\newcommand{\nphys}{Nat. Phys.} 
\newcommand{\na}{New Astron.} 
\newcommand{\nar}{New Astron. Rev.} 
\newcommand{\physrep}{Phys. Rep.} 
\newcommand{\pra}{Phys. Rev. A} 
\newcommand{\prb}{Phys. Rev. B} 
\newcommand{\prc}{Phys. Rev. C} 
\newcommand{\prd}{Phys. Rev. D} 
\newcommand{\pre}{Phys. Rev. E} 
\newcommand{\prx}{Phys. Rev. X} 
\newcommand{\prl}{Phys. Rev. Let.} 
\newcommand{\psj}{Planet. Sci. J.} 
\newcommand{\planss}{Planet. Space Sci.} 
\newcommand{\pnas}{Proc. Natl Acad. Sci. USA} 
\newcommand{\procspie}{Proc. SPIE} 
\newcommand{\pasa}{PASA} 
\newcommand{\pasj}{PASJ} 
\newcommand{\pasp}{PASP} 
\newcommand{\rmxaa}{RMXAA} 
\newcommand{\sci}{Science} 
\newcommand{\sciadv}{Sci. Adv.} 
\newcommand{\solphys}{Sol. Phys.} 
\newcommand{\sovast}{Soviet Ast.} 
\newcommand{\ssr}{Space Sci. Rev.} 
\newcommand{\uni}{Universe} 

%% file: main.bbl
\begin{thebibliography}{90}
\providecommand{\natexlab}[1]{#1}
\providecommand{\url}[1]{\texttt{#1}}
\expandafter\ifx\csname urlstyle\endcsname\relax
  \providecommand{\doi}[1]{doi: #1}\else
  \providecommand{\doi}{doi: \begingroup \urlstyle{rm}\Url}\fi

\bibitem[Afonso et~al.(2026)Afonso, author2, author3, author4, and
  author5]{Afonso2026.SKA}
J.~Afonso et al.
\newblock In \emph{Advancing Astrophysics with the SKA -- II (AASKAII)}. 2026.
\newblock arXiv search: Report number AASKAII/Afonso.1.

\bibitem[{Algera} et~al.(2020){Algera}, {van der Vlugt}, {Hodge}, {Smail},
  {Novak}, {Radcliffe}, {Riechers}, {R{\"o}ttgering}, {Smol{\v{c}}i{\'c}}, and
  {Walter}]{algera20}
H.~S.~B. {Algera} et al.
\newblock \emph{\apj}, 903\penalty0 (2):\penalty0 139, Nov. 2020.
\newblock \doi{10.3847/1538-4357/abb77a}.

\bibitem[Algera et~al.(2026)Algera, author2, author3, author4, and
  author5]{Algera2026.SKA}
H.~S.~B. Algera et al.
\newblock In \emph{Advancing Astrophysics with the SKA -- II (AASKAII)}. 2026.
\newblock arXiv search: Report number AASKAII/Algera.1.

\bibitem[{ALMA Observatory}(2024{\natexlab{a}})]{ALMAWSUFactsheet}
{ALMA Observatory}.
\newblock {ALMA WSU Factsheet and WSU Program documentation}.
\newblock
  \url{https://www.almaobservatory.org/en/alma-development/alma-2030-wideband-sensitivity-upgrade/},
  2024{\natexlab{a}}.
\newblock Accessed 2024-2025.

\bibitem[{ALMA Observatory}(2024{\natexlab{b}})]{ALMAWSUScience}
{ALMA Observatory}.
\newblock {ALMA WSU science case summaries}.
\newblock
  \url{https://www.almaobservatory.org/en/alma-development/alma-2030-wideband-sensitivity-upgrade/alma-2030-wsu-science/},
  2024{\natexlab{b}}.
\newblock ALMA2030 documents and workshop proceedings. Accessed 2024-2025.

\bibitem[{ALMA Observatory}(2024{\natexlab{c}})]{ALMAWSUpage}
{ALMA Observatory}.
\newblock {WSU Program and WSU Science pages (ALMA2030 project pages)}.
\newblock
  \url{https://www.almaobservatory.org/en/alma-development/alma-2030-wideband-sensitivity-upgrade/alma-2030-wsu-science/},
  2024{\natexlab{c}}.
\newblock Accessed 2024-2025.

\bibitem[An et~al.(2026)An, author2, author3, author4, and author5]{An2026.SKA}
F.~X. An et al.
\newblock In \emph{Advancing Astrophysics with the SKA -- II (AASKAII)}. 2026.
\newblock arXiv search: Report number AASKAII/An.4.

\bibitem[{Bacon} et~al.(2024){Bacon}, {Maineiri}, {Randich}, {Cimatti},
  {Kneib}, {Brinchmann}, {Ellis}, {Tolstoi}, {Smiljanic}, {Hill}, {Anderson},
  {Sanchez Saez}, {Opitom}, {Bryson}, {Dierickx}, {Garilli}, {Gonzalez}, {de
  Jong}, {Lee}, {Mieske}, {Otarola}, {Schipani}, {Travouillon}, {Vernet},
  {Bryant}, {Casali}, {Colless}, {Couch}, {Driver}, {Fontana}, {Lehnert},
  {Magrini}, {Montet}, {Pasquini}, {Roth}, {Sanchez-Janssen}, {Steinmetz},
  {Tresse}, {Yeche}, and {Ziegler}]{Bacon2024}
R.~{Bacon} et al.
\newblock In H.~K. {Marshall}, J.~{Spyromilio}, and T.~{Usuda}, editors,
  \emph{Ground-based and Airborne Telescopes X}, volume 13094 of \emph{Society
  of Photo-Optical Instrumentation Engineers (SPIE) Conference Series}, page
  130941O, Aug. 2024.
\newblock \doi{10.1117/12.3018093}.

\bibitem[Bait et~al.(2026)Bait, author2, author3, author4, and
  author5]{Bait2026.SKA}
O.~S. Bait et al.
\newblock In \emph{Advancing Astrophysics with the SKA -- II (AASKAII)}. 2026.
\newblock arXiv search: Report number AASKAII/Bait.1.

\bibitem[Baldi et~al.(2026)Baldi, author2, author3, author4, and
  author5]{Baldi2026.SKA}
R.~D. Baldi et al.
\newblock In \emph{Advancing Astrophysics with the SKA -- II (AASKAII)}. 2026.
\newblock arXiv search: Report number AASKAII/Baldi.1.

\bibitem[{Best} et~al.(2007){Best}, {von der Linden}, {Kauffmann}, {Heckman},
  and {Kaiser}]{Best2007}
P.~N. {Best} et al.
\newblock \emph{\mnras}, 379\penalty0 (3):\penalty0 894--908, Aug. 2007.
\newblock \doi{10.1111/j.1365-2966.2007.11937.x}.

\bibitem[{Best} et~al.(2023){Best}, {Kondapally}, {Williams}, {Cochrane},
  {Duncan}, {Hale}, {Haskell}, {Ma{\l}ek}, {McCheyne}, {Smith}, {Wang},
  {Botteon}, {Bonato}, {Bondi}, {Calistro Rivera}, {Gao}, {G{\"u}rkan},
  {Hardcastle}, {Jarvis}, {Mingo}, {Miraghaei}, {Morabito}, {Nisbet},
  {Prandoni}, {R{\"o}ttgering}, {Sabater}, {Shimwell}, {Tasse}, and {van
  Weeren}]{Best2023}
P.~N. {Best} et al.
\newblock \emph{\mnras}, 523\penalty0 (2):\penalty0 1729--1755, Aug. 2023.
\newblock \doi{10.1093/mnras/stad1308}.

\bibitem[{Bonaldi} et~al.(2023){Bonaldi}, {Hartley}, {Ronconi}, {De Zotti}, and
  {Bonato}]{bonaldi23}
A.~{Bonaldi} et al.
\newblock \emph{\mnras}, 524\penalty0 (1):\penalty0 993--1007, Sept. 2023.
\newblock \doi{10.1093/mnras/stad1913}.

\bibitem[{Bonato} et~al.(2017){Bonato}, {Negrello}, {Mancuso}, {De Zotti},
  {Ciliegi}, {Cai}, {Lapi}, {Massardi}, {Bonaldi}, {Sajina}, and
  et~al.]{bonato17}
M.~{Bonato} et al.
\newblock \emph{\mnras}, 469\penalty0 (2):\penalty0 1912--1923, Aug. 2017.
\newblock \doi{10.1093/mnras/stx974}.

\bibitem[{Bower} et~al.(2006){Bower}, {Benson}, {Malbon}, {Helly}, {Frenk},
  {Baugh}, {Cole}, and {Lacey}]{Bower2006}
R.~G. {Bower} et al.
\newblock \emph{\mnras}, 370\penalty0 (2):\penalty0 645--655, Aug. 2006.
\newblock \doi{10.1111/j.1365-2966.2006.10519.x}.

\bibitem[{Braun} et~al.(2015){Braun}, , , , , , , , , , , , , and {et
  al.}]{Braun2015}
R.~{Braun} et al.
\newblock {SKA1 Science (Level 0) Requirements}.
\newblock Technical Report SKA-TEL-SKO-0000007 Rev 2, SKA Organisation, 2015.
\newblock URL
  \url{https://www.skao.int/sites/default/files/documents/d1_v2-SKA-TEL-SKO-0000007-02-SKA1_Science_Requirements.pdf}.

\bibitem[{Braun} et~al.(2019){Braun}, {Bonaldi}, {Bourke}, {Keane}, and
  {Wagg}]{braun19}
R.~{Braun} et al.
\newblock \emph{arXiv e-prints}, art. arXiv:1912.12699, Dec. 2019.
\newblock \doi{10.48550/arXiv.1912.12699}.

\bibitem[{Carpenter} et~al.(2022){Carpenter}, {Brogan}, {Iono}, {Mroczkowski},
  and {et al.}]{Carpenter2022}
J.~{Carpenter} et al.
\newblock {The ALMA2030 Wideband Sensitivity Upgrade}, Nov 2022.
\newblock ALMA Memo 621.

\bibitem[{Condon}(1992)]{Condon1992}
J.~J. {Condon}.
\newblock \emph{Annual Review of Astronomy and Astrophysics}, 30:\penalty0
  575--611, 1992.
\newblock \doi{10.1146/annurev.aa.30.090192.003043}.

\bibitem[{Coogan} et~al.(2018){Coogan}, {Daddi}, {Sargent}, {Strazzullo},
  {Valentino}, {Gobat}, {Magdis}, {Bethermin}, {Pannella}, {Onodera}, {Liu},
  {Cimatti}, {Dannerbauer}, {Carollo}, {Renzini}, and {Tremou}]{coogan18}
R.~T. {Coogan} et al.
\newblock \emph{\mnras}, 479\penalty0 (1):\penalty0 703--729, Sept. 2018.
\newblock \doi{10.1093/mnras/sty1446}.

\bibitem[{Coogan} et~al.(2023){Coogan}, {Sargent}, {Cibinel}, {Prandoni},
  {Bonaldi}, {Daddi}, and {Franco}]{coogan23}
R.~T. {Coogan} et al.
\newblock \emph{\mnras}, 525\penalty0 (3):\penalty0 3413--3438, Nov. 2023.
\newblock \doi{10.1093/mnras/stad2469}.

\bibitem[{Costantin} et~al.(2023){Costantin}, {P{\'e}rez-Gonz{\'a}lez},
  {Vega-Ferrero}, {Huertas-Company}, {Bisigello}, {Buitrago}, {Bagley},
  {Cleri}, {Cooper}, {Finkelstein}, {Holwerda}, {Kartaltepe}, {Koekemoer},
  {Nelson}, {Papovich}, {Pillepich}, {Pirzkal}, {Tacchella}, and
  {Yung}]{Costantin2023}
L.~{Costantin} et al.
\newblock \emph{\apj}, 946\penalty0 (2):\penalty0 71, Apr. 2023.
\newblock \doi{10.3847/1538-4357/acb926}.

\bibitem[{Croton}(2006)]{Croton2006}
D.~J. {Croton}.
\newblock \emph{\mnras}, 369\penalty0 (4):\penalty0 1808--1812, July 2006.
\newblock \doi{10.1111/j.1365-2966.2006.10429.x}.

\bibitem[{Daddi} et~al.(2017){Daddi}, {Jin}, {Strazzullo}, {Sargent}, {Wang},
  {Ferrari}, {Schinnerer}, {Smol{\v{c}}i{\'c}}, {Calabr{\'o}}, {Coogan},
  {Delhaize}, {Delvecchio}, {Elbaz}, {Gobat}, {Gu}, {Liu}, {Novak}, and
  {Valentino}]{daddi17}
E.~{Daddi} et al.
\newblock \emph{\apjl}, 846\penalty0 (2):\penalty0 L31, Sept. 2017.
\newblock \doi{10.3847/2041-8213/aa8808}.

\bibitem[{Delvecchio} et~al.(2017){Delvecchio}, {Smol{\v{c}}i{\'c}},
  {Zamorani}, {Lagos}, {Berta}, {Delhaize}, {Baran}, {Alexander}, {Rosario},
  {Gonzalez-Perez}, {Ilbert}, {Lacey}, {Le F{\`e}vre}, {Miettinen}, {Aravena},
  {Bondi}, {Carilli}, {Ciliegi}, {Mooley}, {Novak}, {Schinnerer}, {Capak},
  {Civano}, {Fanidakis}, {Herrera Ruiz}, {Karim}, {Laigle}, {Marchesi},
  {McCracken}, {Middleberg}, {Salvato}, and {Tasca}]{Delvecchio2017}
I.~{Delvecchio} et al.
\newblock \emph{\aap}, 602:\penalty0 A3, June 2017.
\newblock \doi{10.1051/0004-6361/201629367}.

\bibitem[{Dickinson} et~al.(2018){Dickinson}, {Ali-Ha{\"\i}moud}, {Barr},
  {Battye}, {Clesse}, {Davis}, {Genova-Santos}, {Harper}, {Ho}, {Jones},
  {Lazarian}, {Marshall}, {Peel}, {Peláez}, {Poletti}, {Prandoni},
  {R{\`a}fols}, {Spinelli}, and {Vidal}]{Dickinson2018}
C.~{Dickinson} et al.
\newblock \emph{New Astronomy Reviews}, 80:\penalty0 1--28, May 2018.
\newblock \doi{10.1016/j.newar.2018.04.001}.

\bibitem[{Duivenvoorden} et~al.(2016){Duivenvoorden}, {Oliver}, {Buat},
  {Darvish}, {Efstathiou}, {Farrah}, {Griffin}, {Hurley}, {Ibar}, {Jarvis},
  {Papadopoulos}, {Sargent}, {Scott}, {Scudder}, {Symeonidis}, {Vaccari},
  {Viero}, and {Wang}]{duivenvoorden16}
S.~{Duivenvoorden} et al.
\newblock \emph{\mnras}, 462\penalty0 (1):\penalty0 277--289, Oct. 2016.
\newblock \doi{10.1093/mnras/stw1466}.

\bibitem[{Dunlop} et~al.(2017){Dunlop}, {McLure}, {Biggs}, {Geach},
  {Micha{\l}owski}, {Ivison}, {Rujopakarn}, {van Kampen}, {Kirkpatrick},
  {Pope}, {Scott}, {Swinbank}, {Targett}, {Aretxaga}, {Austermann}, {Best},
  {Bruce}, {Chapin}, {Charlot}, {Cirasuolo}, {Coppin}, {Ellis}, {Finkelstein},
  {Hayward}, {Hughes}, {Ibar}, {Jagannathan}, {Khochfar}, {Koprowski},
  {Narayanan}, {Nyland}, {Papovich}, {Peacock}, {Rieke}, {Robertson},
  {Vernstrom}, {Werf}, {Wilson}, and {Yun}]{Dunlop2017}
J.~S. {Dunlop} et al.
\newblock \emph{\mnras}, 466\penalty0 (1):\penalty0 861--883, Apr. 2017.
\newblock \doi{10.1093/mnras/stw3088}.

\bibitem[{Erfanianfar} et~al.(2016){Erfanianfar}, {Popesso}, {Finoguenov},
  {Wilman}, {Wuyts}, {Biviano}, {Salvato}, {Mirkazemi}, {Morselli}, {Ziparo},
  {Nandra}, {Lutz}, {Elbaz}, {Dickinson}, {Tanaka}, {Altieri}, {Aussel},
  {Bauer}, {Berta}, {Bielby}, {Brandt}, {Cappelluti}, {Cimatti}, {Cooper},
  {Fadda}, {Ilbert}, {Le Floch}, {Magnelli}, {Mulchaey}, {Nordon}, {Newman},
  {Poglitsch}, and {Pozzi}]{erfanianfar16}
G.~{Erfanianfar} et al.
\newblock \emph{\mnras}, 455\penalty0 (3):\penalty0 2839--2851, Jan. 2016.
\newblock \doi{10.1093/mnras/stv2485}.

\bibitem[{Euclid Collaboration} et~al.(2026){Euclid Collaboration}, {Aussel,
  H.}, {Tereno, I.}, {Schirmer, M.}, {Alguero, G.}, {Altieri, B.}, {Balbinot,
  E.}, {de Boer, T.}, {Casenove, P.}, {Corcho-Caballero, P.}, {Furusawa, H.},
  {Furusawa, J.}, {Hudson, M. J.}, {Jahnke, K.}, {Libet, G.}, {Macias-Perez,
  J.}, {Masoumzadeh, N.}, {Mohr, J. J.}, {Odier, J.}, {Scott, D.}, {Vassallo,
  T.}, {Verdoes Kleijn, G.}, {Zacchei, A.}, {Aghanim, N.}, {Amara, A.},
  {Andreon, S.}, {Auricchio, N.}, {Awan, S.}, {Azzollini, R.}, {Baccigalupi,
  C.}, {Baldi, M.}, {Balestra, A.}, {Bardelli, S.}, {Basset, A.}, {Battaglia,
  P.}, {Belikov, A. N.}, {Bender, R.}, {Biviano, A.}, {Bonchi, A.}, {Bonino,
  D.}, {Branchini, E.}, {Brescia, M.}, {Brinchmann, J.}, {Camera, S.},
  {Ca\~nas-Herrera, G.}, {Capobianco, V.}, {Carbone, C.}, {Cardone, V. F.},
  {Carretero, J.}, {Casas, S.}, {Castander, F. J.}, {Castellano, M.},
  {Castignani, G.}, {Cavuoti, S.}, {Chambers, K. C.}, {Cimatti, A.},
  {Colodro-Conde, C.}, {Congedo, G.}, {Conselice, C. J.}, {Conversi, L.},
  {Copin, Y.}, {Courbin, F.}, {Courtois, H. M.}, {Cropper, M.}, {Cuby, J.-G.},
  {Da Silva, A.}, {da Silva, R.}, {Degaudenzi, H.}, {de Jong, J. T. A.}, {De
  Lucia, G.}, {Di Giorgio, A. M.}, {Dinis, J.}, {Dolding, C.}, {Dole, H.},
  {Douspis, M.}, {Dubath, F.}, {Duncan, C. A. J.}, {Dupac, X.}, {Dusini, S.},
  {Ealet, A.}, {Escoffier, S.}, {Fabricius, M.}, {Farina, M.}, {Farinelli, R.},
  {Faustini, F.}, {Ferriol, S.}, {Fotopoulou, S.}, {Fourmanoit, N.}, {Frailis,
  M.}, {Franceschi, E.}, {Franzetti, P.}, {Galeotta, S.}, {George, K.},
  {Gillard, W.}, {Gillis, B.}, {Giocoli, C.}, {G\'omez-Alvarez, P.},
  {Gracia-Carpio, J.}, {Granett, B. R.}, {Grazian, A.}, {Grupp, F.}, {Guzzo,
  L.}, {Gwyn, S.}, {Haugan, S. V. H.}, {Herent, O.}, {Hoar, J.}, {Hoekstra,
  H.}, {Holliman, M. S.}, {Holmes, W.}, {Hook, I. M.}, {Hormuth, F.},
  {Hornstrup, A.}, {Hudelot, P.}, {Ili\'{}c, S.}, {Jhabvala, M.}, {Joachimi,
  B.}, {Keih\"anen, E.}, {Kermiche, S.}, {Kiessling, A.}, {Kubik, B.},
  {Kuijken, K.}, {K\"ummel, M.}, {Kunz, M.}, {Kurki-Suonio, H.}, {Lahav, O.},
  {Le Boulc'h, Q.}, {Le Brun, A. M. C.}, {Le Mignant, D.}, {Liebing, P.},
  {Ligori, S.}, {Lilje, P. B.}, {Lindholm, V.}, {Lloro, I.}, {Mainetti, G.},
  {Maino, D.}, {Maiorano, E.}, {Mansutti, O.}, {Marcin, S.}, {Marggraf, O.},
  {Markovic, K.}, {Martinelli, M.}, {Martinet, N.}, {Marulli, F.}, {Massey,
  R.}, {Maurogordato, S.}, {McCracken, H. J.}, {Medinaceli, E.}, {Mei, S.},
  {Melchior, M.}, {Mellier, Y.}, {Meneghetti, M.}, {Merlin, E.}, {Meylan, G.},
  {Mora, A.}, {Moresco, M.}, {Morris, P. W.}, {Moscardini, L.}, {Mourre, S.},
  {Nakajima, R.}, {Neissner, C.}, {Nichol, R. C.}, {Niemi, S.-M.},
  {Nightingale, J. W.}, {Nutma, T.}, {Padilla, C.}, {Paltani, S.}, {Pasian,
  F.}, {Peacock, J. A.}, {Pedersen, K.}, {Percival, W. J.}, {Pettorino, V.},
  {Pires, S.}, {Polenta, G.}, {Pollack, J. E.}, {Poncet, M.}, {Popa, L. A.},
  {Pozzetti, L.}, {Racca, G. D.}, {Raison, F.}, {Rebolo, R.}, {Renzi, A.},
  {Rhodes, J.}, {Riccio, G.}, {Rix, H.-W.}, {Romelli, E.}, {Roncarelli, M.},
  {Rossetti, E.}, {Rusholme, B.}, {Saglia, R.}, {Sakr, Z.}, {S\'anchez, A. G.},
  {Sapone, D.}, {Sartoris, B.}, {Sauvage, M.}, {Schewtschenko, J. A.},
  {Schneider, P.}, {Scodeggio, M.}, {Secroun, A.}, {Sefusatti, E.}, {Seidel,
  G.}, {Seiffert, M.}, {Serrano, S.}, {Simon, P.}, {Sirignano, C.}, {Sirri,
  G.}, {Skottfelt, J.}, {Spurio Mancini, A.}, {Stanco, L.}, {Steinwagner, J.},
  {Surace, C.}, {Tallada-Cresp\'{\i}, P.}, {Tavagnacco, D.}, {Taylor, A. N.},
  {Teplitz, H. I.}, {Tessore, N.}, {Toft, S.}, {Toledo-Moreo, R.},
  {Torradeflot, F.}, {Tsyganov, A.}, {Tutusaus, I.}, {Valentijn, E. A.},
  {Valenziano, L.}, {Valiviita, J.}, {Veropalumbo, A.}, {Wang, Y.}, {Weller,
  J.}, {Williams, O. R.}, {Zamorani, G.}, {Zerbi, F. M.}, {Zucca, E.},
  {Allevato, V.}, {Ballardini, M.}, {Blake, R. P.}, {Bolzonella, M.}, {Bozzo,
  E.}, {Burigana, C.}, {Cabanac, R.}, {Calabrese, M.}, {Cappi, A.}, {Di
  Ferdinando, D.}, {Escartin Vigo, J. A.}, {Gabarra, L.}, {Hartley, W. G.},
  {Huertas-Company, M.}, {Mart\'{\i}n-Fleitas, J.}, {Matthew, S.}, {Maturi,
  M.}, {Mauri, N.}, {Metcalf, R. B.}, {Pezzotta, A.}, {P\"ontinen, M.},
  {Porciani, C.}, {Risso, I.}, {Scottez, V.}, {Sereno, M.}, {Tenti, M.}, {Viel,
  M.}, {Wiesmann, M.}, {Akrami, Y.}, {Alvi, S.}, {Andika, I. T.}, {Anselmi,
  S.}, {Archidiacono, M.}, {Atrio-Barandela, F.}, {Avila, S.}, {Bergamini, P.},
  {Bertacca, D.}, {Bethermin, M.}, {Bisigello, L.}, {Blanchard, A.}, {Blot,
  L.}, {B\"ohringer, H.}, {Borgani, S.}, {Borlaff, A. S.}, {Brown, M. L.},
  {Bruton, S.}, {Buitrago, F.}, {Calabro, A.}, {Calderone, G.}, {Camacho
  Quevedo, B.}, {Caro, F.}, {Carvalho, C. S.}, {Castro, T.}, {Charles, Y.},
  {Cogato, F.}, {Conseil, S.}, {Cooray, A. R.}, {Costanzi, M.}, {Cucciati, O.},
  {Davini, S.}, {De Paolis, F.}, {Desprez, G.}, {D\'{\i}az-S\'anchez, A.},
  {Diaz, J. J.}, {Di Domizio, S.}, {Diego, J. M.}, {Dimauro, P.}, {Duc, P.-A.},
  {Enia, A.}, {Fang, Y.}, {Ferguson, A. M. N.}, {Ferrari, A. G.}, {Finoguenov,
  A.}, {Fontana, A.}, {Fontanot, F.}, {Franco, A.}, {Garc\'{\i}a-Bellido, J.},
  {Gasparetto, T.}, {Gavazzi, R.}, {Gaztanaga, E.}, {Giacomini, F.}, {Gianotti,
  F.}, {Gonzalez, A. H.}, {Gozaliasl, G.}, {Gruppuso, A.}, {Guidi, M.},
  {Gutierrez, C. M.}, {Hall, A.}, {Hern\'andez-Monteagudo, C.}, {Hildebrandt,
  H.}, {Hjorth, J.}, {Jacobson, J.}, {Joudaki, S.}, {Kajava, J. J. E.}, {Kang,
  Y.}, {Kansal, V.}, {Karagiannis, D.}, {Kiiveri, K.}, {Kirkpatrick, C. C.},
  {Kruk, S.}, {Lacasa, F.}, {Laigle, C.}, {Lattanzi, M.}, {Le Brun, V.}, {Le
  Graet, J.}, {Legrand, L.}, {Lembo, M.}, {Lepori, F.}, {Leroy, G.}, {Lesci, G.
  F.}, {Lesgourgues, J.}, {Leuzzi, L.}, {Liaudat, T. I.}, {Loureiro, A.},
  {Magliocchetti, M.}, {Magnier, E. A.}, {Mancini, C.}, {Mannucci, F.}, {Maoli,
  R.}, {Martins, C. J. A. P.}, {Maurin, L.}, {McPartland, C. J. R.}, {Melin,
  J.-B.}, {Migliaccio, M.}, {Miluzio, M.}, {Monaco, P.}, {Montoro, A.},
  {Moretti, C.}, {Morgante, G.}, {Murray, C.}, {Nadathur, S.}, {Naidoo, K.},
  {Navarro-Alsina, A.}, {Nesseris, S.}, {Nicastro, L.}, {Oguri, M.},
  {Passalacqua, F.}, {Paterson, K.}, {Patrizii, L.}, {Pisani, A.}, {Potter,
  D.}, {Quai, S.}, {Radovich, M.}, {Reimberg, P.}, {Rocci, P.-F.}, {Rodighiero,
  G.}, {Rollins, R. P.}, {Sacquegna, S.}, {Sahl\'en, M.}, {Sanders, D. B.},
  {Sarpa, E.}, {Scarlata, C.}, {Schaye, J.}, {Schneider, A.}, {Schultheis, M.},
  {Sciotti, D.}, {Scognamiglio, D.}, {Sellentin, E.}, {Shankar, F.}, {Smith, L.
  C.}, {Soubrie, E.}, {Stanford, S. A.}, {Tanidis, K.}, {Tao, C.}, {Testera,
  G.}, {Tewes, M.}, {Teyssier, R.}, {Tosi, S.}, {Troja, A.}, {Tucci, M.},
  {Valieri, C.}, {Venhola, A.}, {Vergani, D.}, {Vernizzi, F.}, {Verza, G.},
  {Vielzeuf, P.}, {Walton, N. A.}, {Weaver, J. R.}, {Wilde, J.}, and {Zalesky,
  L.}]{Ausssel2025}
{Euclid Collaboration} et al.
\newblock \emph{A\&A}, 2026.
\newblock \doi{10.1051/0004-6361/202554610}.
\newblock URL \url{https://doi.org/10.1051/0004-6361/202554610}.

\bibitem[{Gaspari} et~al.(2019){Gaspari}, {Eckert}, {Ettori}, {Tozzi},
  {Bassini}, {Rasia}, {Brighenti}, {Sun}, {Borgani}, {Johnson}, {Tremblay},
  {Stone}, {Temi}, {Yang}, {Tombesi}, and {Cappi}]{Gaspari2019}
M.~{Gaspari} et al.
\newblock \emph{\apj}, 884\penalty0 (2):\penalty0 169, Oct. 2019.
\newblock \doi{10.3847/1538-4357/ab3c5d}.

\bibitem[Hardcastle et~al.(2026)Hardcastle, author2, author3, author4, and
  author5]{Hardcastle2026.SKA}
M.~J. Hardcastle et al.
\newblock In \emph{Advancing Astrophysics with the SKA -- II (AASKAII)}. 2026.
\newblock arXiv search: Report number AASKAII/Hardcastle.1.

\bibitem[{Hassani} et~al.(2022){Hassani}, {Tabatabaei}, {Hughes}, {Chastenet},
  {McLeod}, {Schinnerer}, and {Nasiri}]{Hassani}
H.~{Hassani} et al.
\newblock \emph{\mnras}, 510\penalty0 (1):\penalty0 11--31, Feb. 2022.
\newblock \doi{10.1093/mnras/stab3202}.

\bibitem[{Heckman}(1983)]{Heckman1983}
T.~M. {Heckman}.
\newblock \emph{\apj}, 268:\penalty0 628--631, May 1983.
\newblock \doi{10.1086/160984}.

\bibitem[{Heckman} and {Best}(2014)]{Heckman2014}
T.~M. {Heckman} and P.~N. {Best}.
\newblock \emph{\araa}, 52:\penalty0 589--660, Aug. 2014.
\newblock \doi{10.1146/annurev-astro-081913-035722}.

\bibitem[{Hopkins} et~al.(2025){Hopkins}, {Kapinska}, {Marvil}, {Vernstrom},
  {Collier}, {Norris}, {Gordon}, {Duchesne}, {Rudnick}, {Gupta}, {Carretti},
  {Anderson}, {Dai}, {G{\"u}rkan}, {Parkinson}, {Prandoni}, {Riggi}, {Shekhar
  Saraf}, {Ma}, {Filipovi{\'c}}, {Umana}, {Bahr-Kalus}, {Koribalski}, {Lenc},
  {Ingallinera}, {Afonso}, {Ahmad}, {Ahmed}, {Alexander}, {Andernach},
  {Asorey}, {Battisti}, {Bilicki}, {Botteon}, {Brown}, {Br{\"u}ggen}, {Cowley},
  {Dage}, {Hale}, {Hardcastle}, {Kothes}, {Lazarevi{\'c}}, {Lin}, {Luken},
  {Moss}, {Prathap}, {ur Rahman}, {Reiprich}, {Riseley}, {Salvato}, {Seymour},
  {Shabala}, {Smith}, {Vaccari}, {van Loon}, {Wong}, {Zainal Alsaberi},
  {Asher}, {Ball}, {Barbosa}, {Biava}, {Bradley}, {Carvajal}, {Crawford},
  {Galvin}, {Huynh}, {Leahy}, {Matute}, {Moss}, {Pappalardo}, {Smeaton},
  {Velovi{\'c}}, and {Zafar}]{Hopkins2025}
A.~{Hopkins} et al.
\newblock \emph{\pasa}, 42:\penalty0 e071, May 2025.
\newblock \doi{10.1017/pasa.2025.10042}.

\bibitem[{Hopkins} et~al.(2012){Hopkins}, {Kere{\v{s}}}, {Murray}, {Quataert},
  and {Hernquist}]{Hopkins2012}
P.~F. {Hopkins} et al.
\newblock \emph{\mnras}, 427\penalty0 (2):\penalty0 968--978, Dec. 2012.
\newblock \doi{10.1111/j.1365-2966.2012.21981.x}.

\bibitem[{Jarvis} et~al.(2016){Jarvis}, {Taylor}, {Agudo}, {Allison}, {Deane},
  {Frank}, {Gupta}, {Heywood}, {Maddox}, {McAlpine}, {Santos}, {Scaife},
  {Vaccari}, {Zwart}, {Adams}, {Bacon}, {Baker}, {Bassett}, {Best}, {Beswick},
  {Blyth}, {Brown}, {Bruggen}, {Cluver}, {Colafrancesco}, {Cotter}, {Cress},
  {Dav{\'e}}, {Ferrari}, {Hardcastle}, {Hale}, {Harrison}, {Hatfield},
  {Klockner}, {Kolwa}, {Malefahlo}, {Marubini}, {Mauch}, {Moodley}, {Morganti},
  {Norris}, {Peters}, {Prandoni}, {Prescott}, {Oliver}, {Oozeer}, {Rottgering},
  {Seymour}, {Simpson}, {Smirnov}, and {Smith}]{Jarvis2016}
M.~{Jarvis} et al.
\newblock In \emph{MeerKAT Science: On the Pathway to the SKA}, page~6, Jan.
  2016.
\newblock \doi{10.22323/1.277.0006}.

\bibitem[{Jim{\'e}nez-Andrade} et~al.(2019){Jim{\'e}nez-Andrade}, {Magnelli},
  {Karim}, {Zamorani}, {Bondi}, {Schinnerer}, {Sargent}, {Romano-D{\'\i}az},
  {Novak}, {Lang}, {Bertoldi}, {Vardoulaki}, {Toft}, {Smol{\v{c}}i{\'c}},
  {Harrington}, {Leslie}, {Delhaize}, {Liu}, {Karoumpis}, {Kartaltepe}, and
  {Koekemoer}]{jimenez-andrade19}
E.~F. {Jim{\'e}nez-Andrade} et al.
\newblock \emph{\aap}, 625:\penalty0 A114, May 2019.
\newblock \doi{10.1051/0004-6361/201935178}.

\bibitem[{Jim{\'e}nez-Andrade} et~al.(2024){Jim{\'e}nez-Andrade}, {Murphy},
  {Momjian}, {Condon}, {Chary}, {Taylor}, and {Dickinson}]{jimenez-andrade24}
E.~F. {Jim{\'e}nez-Andrade} et al.
\newblock \emph{\apj}, 972\penalty0 (1):\penalty0 89, Sept. 2024.
\newblock \doi{10.3847/1538-4357/ad5b5c}.

\bibitem[{Keel}(1984)]{Keel1984}
W.~C. {Keel}.
\newblock \emph{\apj}, 282:\penalty0 75--84, July 1984.
\newblock \doi{10.1086/162177}.

\bibitem[Kondapally et~al.(2026)Kondapally, author2, author3, author4, and
  author5]{Kondapally2026.SKA}
R.~Kondapally et al.
\newblock In \emph{Advancing Astrophysics with the SKA -- II (AASKAII)}. 2026.
\newblock arXiv search: Report number AASKAII/Kondapally.1.

\bibitem[{Kormendy} and {Ho}(2013)]{Kormendy2013}
J.~{Kormendy} and L.~C. {Ho}.
\newblock \emph{\araa}, 51\penalty0 (1):\penalty0 511--653, Aug. 2013.
\newblock \doi{10.1146/annurev-astro-082708-101811}.

\bibitem[{Kratzer} and {Richards}(2015)]{Kratzer15}
R.~M. {Kratzer} and G.~T. {Richards}.
\newblock \emph{\aj}, 149\penalty0 (2):\penalty0 61, Feb. 2015.
\newblock \doi{10.1088/0004-6256/149/2/61}.

\bibitem[{Lal} et~al.(2011){Lal}, {Shastri}, and {Gabuzda}]{LalShastri2011}
D.~V. {Lal}, P.~{Shastri}, and D.~C. {Gabuzda}.
\newblock \emph{\apj}, 731\penalty0 (1):\penalty0 68, Apr. 2011.
\newblock \doi{10.1088/0004-637X/731/1/68}.

\bibitem[{Lal} et~al.(2025){Lal}, {Taylor}, {Sekhar}, {Ishwara-Chandra},
  {Dutta}, and {Kolwa}]{Lal2025}
D.~V. {Lal} et al.
\newblock \emph{\apj}, 991\penalty0 (1):\penalty0 9, Sept. 2025.
\newblock \doi{10.3847/1538-4357/adf6dc}.

\bibitem[{Leslie} et~al.(2020){Leslie}, {Schinnerer}, {Liu}, {Magnelli},
  {Algera}, {Karim}, {Davidzon}, {Gozaliasl}, {Jim{\'e}nez-Andrade}, {Lang},
  {Sargent}, {Novak}, {Groves}, {Smol{\v{c}}i{\'c}}, {Zamorani}, {Vaccari},
  {Battisti}, {Vardoulaki}, {Peng}, and {Kartaltepe}]{leslie20}
S.~K. {Leslie} et al.
\newblock \emph{\apj}, 899\penalty0 (1):\penalty0 58, Aug. 2020.
\newblock \doi{10.3847/1538-4357/aba044}.

\bibitem[Maccagni et~al.(2026)Maccagni, author2, author3, author4, and
  author5]{Maccagni2026.SKA}
F.~M. Maccagni et al.
\newblock In \emph{Advancing Astrophysics with the SKA -- II (AASKAII)}. 2026.
\newblock arXiv search: Report number AASKAII/Maccagni.1.

\bibitem[{Macfarlane} et~al.(2021){Macfarlane}, {Best}, {Sabater},
  {G{\"u}rkan}, {Jarvis}, {R{\"o}ttgering}, {Baldi}, {Calistro Rivera},
  {Duncan}, {Morabito}, {Prandoni}, and {Retana-Montenegro}]{Macfarlane21}
C.~{Macfarlane} et al.
\newblock \emph{\mnras}, 506\penalty0 (4):\penalty0 5888--5907, Oct. 2021.
\newblock \doi{10.1093/mnras/stab1998}.

\bibitem[{Magliocchetti}(2022)]{Magliocchetti2022}
M.~{Magliocchetti}.
\newblock \emph{\aapr}, 30\penalty0 (1):\penalty0 6, Dec. 2022.
\newblock \doi{10.1007/s00159-022-00142-1}.

\bibitem[{Magliocchetti} et~al.(2017){Magliocchetti}, {Popesso}, {Brusa},
  {Salvato}, {Laigle}, {McCracken}, and {Ilbert}]{Magliocchetti17}
M.~{Magliocchetti} et al.
\newblock \emph{\mnras}, 464\penalty0 (3):\penalty0 3271--3280, Jan. 2017.
\newblock \doi{10.1093/mnras/stw2541}.

\bibitem[{Mazoochi} et~al.(2026){Mazoochi}, {Tabatabaei}, {Barnes}, {Colzi},
  {Garc{\'\i}a}, {Henkel}, {Hu}, {Longmore}, {Mart{\'\i}n},
  {S{\'a}nchez-Monge}, {Rivilla}, {Schmiedeke}, {Ott}, {Walker}, {Wang},
  {Williams}, and {Zhang}]{Mazoochi}
F.~{Mazoochi} et al.
\newblock \emph{\apj}, 997\penalty0 (1):\penalty0 31, Jan. 2026.
\newblock \doi{10.3847/1538-4357/ae1b93}.

\bibitem[{Mazzolari} et~al.(2024){Mazzolari}, {Gilli}, {Brusa}, {Mignoli},
  {Vito}, {Prandoni}, {Marchesi}, {Chiaberge}, {Lanzuisi}, {D'Amato},
  {Comastri}, {Vignali}, {Iwasawa}, and {Norman}]{Mazzolari2024}
G.~{Mazzolari} et al.
\newblock \emph{\aap}, 687:\penalty0 A120, July 2024.
\newblock \doi{10.1051/0004-6361/202348072}.

\bibitem[{McKay} et~al.(2025){McKay}, {Subrahmanyan}, {Chippendale}, {Bolli},
  {Kyriakou}, {Dunning}, and {Ekers}]{mckay25}
L.~{McKay} et al.
\newblock \emph{arXiv e-prints}, art. arXiv:2509.11846, Sept. 2025.
\newblock \doi{10.48550/arXiv.2509.11846}.

\bibitem[Moldon et~al.(2026)Moldon, author2, author3, author4, and
  author5]{Moldon2026.SKA}
J.~Moldon et al.
\newblock In \emph{Advancing Astrophysics with the SKA -- II (AASKAII)}. 2026.
\newblock arXiv search: Report number AASKAII/Moldon.1.

\bibitem[{Morselli} et~al.(2019){Morselli}, {Popesso}, {Cibinel}, {Oesch},
  {Montes}, {Atek}, {Illingworth}, and {Holden}]{morselli19}
L.~{Morselli} et al.
\newblock \emph{\aap}, 626:\penalty0 A61, June 2019.
\newblock \doi{10.1051/0004-6361/201834559}.

\bibitem[{Mundell} et~al.(2009){Mundell}, {Ferruit}, {Nagar}, and
  {Wilson}]{Mundell2009}
C.~G. {Mundell}, P.~{Ferruit}, N.~{Nagar}, and A.~S. {Wilson}.
\newblock \emph{\apj}, 703\penalty0 (1):\penalty0 802--815, Sept. 2009.
\newblock \doi{10.1088/0004-637X/703/1/802}.

\bibitem[{Murphy} et~al.(2015){Murphy}, {Sargent}, {Beswick}, {Dickinson},
  {Heywood}, {Hunt}, {Huynh}, {Jarvis}, {Karim}, {Krause}, and
  et~al.]{Murphy15}
E.~{Murphy} et al.
\newblock In \emph{Advancing Astrophysics with the Square Kilometre Array
  (AASKA14)}, page~85, Apr. 2015.
\newblock \doi{10.22323/1.215.0085}.

\bibitem[{Murphy} et~al.(2011){Murphy}, {Condon}, {Schinnerer}, {Kennicutt},
  {Jahnke}, {Armus}, {Calzetti}, {Daddi}, {Helou}, {Hinz}, {Johnson}, {Dale},
  {Engelbracht}, {Galametz}, {Bendo}, {Appleton}, {Meyer}, {Roussel}, {Smith},
  {Bolatto}, {Crocker}, {Croxall}, {Dumas}, {Groves}, {Hunt}, {Koda}, {Kregel},
  {Meidt}, {Mizusawa}, {Ott}, {Sandstrom}, {Sauvage}, {Srinivasan}, {Walter},
  {Warren}, {Wilson}, {Wolfire}, and {Zibetti}]{Murphy2011}
E.~J. {Murphy} et al.
\newblock \emph{The Astrophysical Journal}, 737\penalty0 (2):\penalty0 67, Aug
  2011.
\newblock \doi{10.1088/0004-637X/737/2/67}.

\bibitem[{Muxlow} et~al.(2020){Muxlow}, {Thomson}, {Radcliffe}, {Wrigley},
  {Beswick}, {Smail}, {McHardy}, {Garrington}, {Ivison}, {Jarvis}, {Prandoni},
  {Bondi}, {Guidetti}, {Argo}, {Bacon}, {Best}, {Biggs}, {Chapman}, {Coppin},
  {Chen}, {Garratt}, {Garrett}, {Ibar}, {Kneib}, {Knudsen}, {Koopmans},
  {Morabito}, {Murphy}, {Njeri}, {Pearson}, {P{\'e}rez-Torres}, {Richards},
  {R{\"o}ttgering}, {Sargent}, {Serjeant}, {Simpson}, {Simpson}, {Swinbank},
  {Varenius}, and {Venturi}]{muxlow20}
T.~W.~B. {Muxlow} et al.
\newblock \emph{\mnras}, 495\penalty0 (1):\penalty0 1188--1208, June 2020.
\newblock \doi{10.1093/mnras/staa1279}.

\bibitem[{Nagar} et~al.(2002){Nagar}, {Falcke}, {Wilson}, and
  {Ulvestad}]{Nagar2002}
N.~M. {Nagar}, H.~{Falcke}, A.~S. {Wilson}, and J.~S. {Ulvestad}.
\newblock \emph{\aap}, 392:\penalty0 53--82, Sept. 2002.
\newblock \doi{10.1051/0004-6361:20020874}.

\bibitem[{Nagar} et~al.(2005){Nagar}, {Falcke}, and {Wilson}]{Nagar2005}
N.~M. {Nagar}, H.~{Falcke}, and A.~S. {Wilson}.
\newblock \emph{\aap}, 435\penalty0 (2):\penalty0 521--543, May 2005.
\newblock \doi{10.1051/0004-6361:20042277}.

\bibitem[{Nelson} et~al.(2019){Nelson}, {Tadaki}, {Tacconi}, {Lutz},
  {F{\"o}rster Schreiber}, {Cibinel}, {Wuyts}, {Lang}, {Leja}, {Montes},
  {Oesch}, {Belli}, {Davies}, {Davies}, {Genzel}, {Lippa}, {Price},
  {{\"U}bler}, and {Wisnioski}]{nelson19}
E.~J. {Nelson} et al.
\newblock \emph{\apj}, 870\penalty0 (2):\penalty0 130, Jan. 2019.
\newblock \doi{10.3847/1538-4357/aaf38a}.

\bibitem[{Norris} et~al.(2011){Norris}, {Hopkins}, {Afonso}, {Brown}, {Condon},
  {Dunne}, {Feain}, {Hollow}, {Jarvis}, {Johnston-Hollitt}, {Lenc},
  {Middelberg}, {Padovani}, {Prandoni}, {Rudnick}, {Seymour}, {Umana},
  {Andernach}, {Alexander}, {Appleton}, {Bacon}, {Banfield}, {Becker}, {Brown},
  {Ciliegi}, {Jackson}, {Eales}, {Edge}, {Gaensler}, {Giovannini}, {Hales},
  {Hancock}, {Huynh}, {Ibar}, {Ivison}, {Kennicutt}, {Kimball}, {Koekemoer},
  {Koribalski}, {L{\'o}pez-S{\'a}nchez}, {Mao}, {Murphy}, {Messias},
  {Pimbblet}, {Raccanelli}, {Randall}, {Reiprich}, {Roseboom},
  {R{\"o}ttgering}, {Saikia}, {Sharp}, {Slee}, {Smail}, {Thompson}, {Urquhart},
  {Wall}, and {Zhao}]{Norris2011}
R.~P. {Norris} et al.
\newblock \emph{\pasa}, 28\penalty0 (3):\penalty0 215--248, Aug. 2011.
\newblock \doi{10.1071/AS11021}.

\bibitem[{Norris} et~al.(2021){Norris}, {Marvil}, {Collier}, {Kapi{\'n}ska},
  {O'Brien}, {Rudnick}, {Andernach}, {Asorey}, {Brown}, {Br{\"u}ggen},
  {Crawford}, {English}, {Rahman}, {Filipovi{\'c}}, {Gordon}, {G{\"u}rkan},
  {Hale}, {Hopkins}, {Huynh}, {HyeongHan}, {James Jee}, {Koribalski}, {Lenc},
  {Luken}, {Parkinson}, {Prandoni}, {Raja}, {Reiprich}, {Riseley}, {Shabala},
  {Sheil}, {Vernstrom}, {Whiting}, {Allison}, {Anderson}, {Ball}, {Bell},
  {Bunton}, {Galvin}, {Gupta}, {Hotan}, {Jacka}, {Macgregor}, {Mahony}, {Maio},
  {Moss}, {Pandey-Pommier}, and {Voronkov}]{Norris2021}
R.~P. {Norris} et al.
\newblock \emph{\pasa}, 38:\penalty0 e046, Sept. 2021.
\newblock \doi{10.1017/pasa.2021.42}.

\bibitem[Panessa et~al.(2026)Panessa, author2, author3, author4, and
  author5]{Panessa2026.SKA}
F.~Panessa et al.
\newblock In \emph{Advancing Astrophysics with the SKA -- II (AASKAII)}. 2026.
\newblock arXiv search: Report number AASKAII/Panessa.1.

\bibitem[{Prandoni} and {Seymour}(2015)]{Prandoni2015}
I.~{Prandoni} and N.~{Seymour}.
\newblock In \emph{{Advancing Astrophysics with the Square Kilometre Array
  (AASKA14)}}, volume 067 of \emph{PoS(AASKA14)}, Jan 2015.
\newblock \doi{10.22323/1.215.0067}.

\bibitem[{Prandoni} et~al.(2024){Prandoni}, {Sargent}, {Adams}, {Catinella},
  {Bonaldi}, {Harrison}, {Mainieri}, {Morganti}, {Best}, {Magliocchetti}, and
  {Zwaan}]{Prandoni2024}
I.~{Prandoni} et al.
\newblock \emph{The Messenger}, 193:\penalty0 14--19, Sep 2024.
\newblock \doi{10.18727/0722-6691/5361}.

\bibitem[{Radcliffe} et~al.(2021{\natexlab{a}}){Radcliffe}, {Barthel},
  {Garrett}, {Beswick}, {Thomson}, and {Muxlow}]{Radcliffe2021b}
J.~F. {Radcliffe} et al.
\newblock \emph{\aap}, 649:\penalty0 L9, May 2021{\natexlab{a}}.
\newblock \doi{10.1051/0004-6361/202140791}.

\bibitem[{Radcliffe} et~al.(2021{\natexlab{b}}){Radcliffe}, {Barthel},
  {Thomson}, {Garrett}, {Beswick}, and {Muxlow}]{Radcliffe2021a}
J.~F. {Radcliffe} et al.
\newblock \emph{\aap}, 649:\penalty0 A27, May 2021{\natexlab{b}}.
\newblock \doi{10.1051/0004-6361/202038591}.

\bibitem[{Rhodes} et~al.(2017){Rhodes}, {Nichol}, {Aubourg}, {Bean},
  {Boutigny}, {Bremer}, {Capak}, {Cardone}, {Carry}, {Conselice}, {Connolly},
  {Cuillandre}, {Hatch}, {Helou}, {Hemmati}, {Hildebrandt}, {Hlo{\v{z}}ek},
  {Jones}, {Kahn}, {Kiessling}, {Kitching}, {Lupton}, {Mandelbaum}, {Markovic},
  {Marshall}, {Massey}, {Maughan}, {Melchior}, {Mellier}, {Newman},
  {Robertson}, {Sauvage}, {Schrabback}, {Smith}, {Strauss}, {Taylor}, and {Von
  Der Linden}]{Rhodes2017}
J.~{Rhodes} et al.
\newblock \emph{\apjs}, 233\penalty0 (2):\penalty0 21, Dec. 2017.
\newblock \doi{10.3847/1538-4365/aa96b0}.

\bibitem[{Sabater} et~al.(2019){Sabater}, {Best}, {Hardcastle}, {Shimwell},
  {Tasse}, {Williams}, {Br{\"u}ggen}, {Cochrane}, {Croston}, {de Gasperin},
  {Duncan}, {G{\"u}rkan}, {Mechev}, {Morabito}, {Prandoni}, {R{\"o}ttgering},
  {Smith}, {Harwood}, {Mingo}, {Mooney}, and {Saxena}]{Sabater2019}
J.~{Sabater} et al.
\newblock \emph{\aap}, 622:\penalty0 A17, Feb. 2019.
\newblock \doi{10.1051/0004-6361/201833883}.

\bibitem[{Shuntov} et~al.(2022){Shuntov}, {McCracken}, {Gavazzi}, {Laigle},
  {Weaver}, {Davidzon}, {Ilbert}, {Kauffmann}, {Faisst}, {Dubois}, {Koekemoer},
  {Moneti}, {Milvang-Jensen}, {Mobasher}, {Sanders}, and {Toft}]{Shuntov2022}
M.~{Shuntov} et al.
\newblock \emph{\aap}, 664:\penalty0 A61, Aug. 2022.
\newblock \doi{10.1051/0004-6361/202243136}.

\bibitem[{Silk} and {Mamon}(2012)]{Silk2012}
J.~{Silk} and G.~A. {Mamon}.
\newblock \emph{Research in Astronomy and Astrophysics}, 12\penalty0
  (8):\penalty0 917--946, Aug. 2012.
\newblock \doi{10.1088/1674-4527/12/8/004}.

\bibitem[{Somerville} and {Dav{\'e}}(2015)]{Somerville2015}
R.~S. {Somerville} and R.~{Dav{\'e}}.
\newblock \emph{\araa}, 53:\penalty0 51--113, Aug. 2015.
\newblock \doi{10.1146/annurev-astro-082812-140951}.

\bibitem[{Tabatabaei} et~al.(2024){Tabatabaei}, {Ghasemi-Nodehi}, {Sargent},
  {Murphy}, {Schinnerer}, {Bonaldi}, and {SKA ISM/IGM Focus Group}]{taba_24}
F.~{Tabatabaei} et al.
\newblock In F.~{Tabatabaei}, B.~{Barbuy}, and Y.-S. {Ting}, editors,
  \emph{Early Disk-Galaxy Formation from JWST to the Milky Way}, volume 377 of
  \emph{IAU Symposium}, pages 43--47, Jan. 2024.
\newblock \doi{10.1017/S1743921323001680}.

\bibitem[{Tabatabaei} et~al.(2025){Tabatabaei}, {Khademi}, {Jarvis}, {Taylor},
  {Whittam}, {An}, {Javadi}, {Murphy}, and {Vaccari}]{taba_25}
F.~{Tabatabaei} et al.
\newblock \emph{\apj}, 989\penalty0 (1):\penalty0 44, Aug. 2025.
\newblock \doi{10.3847/1538-4357/ade233}.

\bibitem[{Tabatabaei} et~al.(2017){Tabatabaei}, {Schinnerer}, {Krause},
  {Dumas}, {Meidt}, {Damas-Segovia}, {Beck}, {Murphy}, {Mulcahy}, {Groves},
  {Bolatto}, {Dale}, {Galametz}, {Sandstrom}, {Boquien}, {Calzetti},
  {Kennicutt}, {Hunt}, {De Looze}, and {Pellegrini}]{taba_17}
F.~S. {Tabatabaei} et al.
\newblock \emph{\apj}, 836\penalty0 (2):\penalty0 185, Feb. 2017.
\newblock \doi{10.3847/1538-4357/836/2/185}.

\bibitem[{Tabatabaei} et~al.(2018){Tabatabaei}, {Minguez}, {Prieto}, and
  {Fern{\'a}ndez-Ontiveros}]{taba_18}
F.~S. {Tabatabaei}, P.~{Minguez}, M.~A. {Prieto}, and J.~A.
  {Fern{\'a}ndez-Ontiveros}.
\newblock \emph{Nature Astronomy}, 2:\penalty0 83--89, Nov. 2018.
\newblock \doi{10.1038/s41550-017-0298-7}.

\bibitem[Tabatabaei et~al.(2026)Tabatabaei, author2, author3, author4, and
  author5]{Tabatabaei2026.SKA}
F.~S. Tabatabaei et al.
\newblock In \emph{Advancing Astrophysics with the SKA -- II (AASKAII)}. 2026.
\newblock arXiv search: Report number AASKAII/Tabatabaei.1.

\bibitem[{Tacconi} et~al.(2020){Tacconi}, {Genzel}, and
  {Sternberg}]{Tacconi2020}
L.~J. {Tacconi}, R.~{Genzel}, and A.~{Sternberg}.
\newblock \emph{\araa}, 58:\penalty0 157--203, Aug. 2020.
\newblock \doi{10.1146/annurev-astro-082812-141034}.

\bibitem[{Thomson} et~al.(2019){Thomson}, {Smail}, {Swinbank}, {Simpson},
  {Arumugam}, {Stach}, {Murphy}, {Rujopakarn}, {Almaini}, {An}, {Blain},
  {Chen}, {Cooke}, {Dudzevi{\v{c}}i{\={u}}t{\.{e}}}, {Edge}, {Farrah},
  {Gullberg}, {Hartley}, {Ibar}, {Maltby}, {Micha{\l}owski}, {Simpson}, {van
  der Werf}, and {Wardlow}]{Thomson19}
A.~P. {Thomson} et al.
\newblock \emph{\apj}, 883\penalty0 (2):\penalty0 204, Oct. 2019.
\newblock \doi{10.3847/1538-4357/ab32e7}.

\bibitem[{Urry} and {Padovani}(1995)]{Urry95}
C.~M. {Urry} and P.~{Padovani}.
\newblock \emph{\pasp}, 107:\penalty0 803, Sept. 1995.
\newblock \doi{10.1086/133630}.

\bibitem[{van Dyk} and {Ho}(1998)]{vanDyk1998}
S.~D. {van Dyk} and L.~C. {Ho}.
\newblock In J.~A. {Zensus}, G.~B. {Taylor}, and J.~M. {Wrobel}, editors,
  \emph{IAU Colloquium 164: Radio Emission from Galactic and Extragalactic
  Compact Sources}, volume 144 of \emph{Astronomical Society of the Pacific
  Conference Series}, page 205, Jan. 1998.

\bibitem[{Vito} et~al.(2018){Vito}, {Brandt}, {Yang}, {Gilli}, {Luo},
  {Vignali}, {Xue}, {Comastri}, {Koekemoer}, {Lehmer}, {Liu}, {Paolillo},
  {Ranalli}, {Schneider}, {Shemmer}, {Volonteri}, and {Wang}]{Vito2018}
F.~{Vito} et al.
\newblock \emph{\mnras}, 473\penalty0 (2):\penalty0 2378--2406, Jan. 2018.
\newblock \doi{10.1093/mnras/stx2486}.

\bibitem[{Weaver} et~al.(2023){Weaver}, {Davidzon}, {Toft}, {Ilbert},
  {McCracken}, {Gould}, {Jespersen}, {Steinhardt}, {Lagos}, {Capak}, and
  et~al.]{weaver23}
J.~R. {Weaver} et al.
\newblock \emph{\aap}, 677:\penalty0 A184, Sept. 2023.
\newblock \doi{10.1051/0004-6361/202245581}.

\bibitem[{Wechsler} and {Tinker}(2018)]{Wechsler2018}
R.~H. {Wechsler} and J.~L. {Tinker}.
\newblock \emph{\araa}, 56:\penalty0 435--487, Sept. 2018.
\newblock \doi{10.1146/annurev-astro-081817-051756}.

\bibitem[{Whittam} et~al.(2022){Whittam}, {Jarvis}, {Hale}, {Prescott},
  {Morabito}, {Heywood}, {Adams}, {Afonso}, {An}, {Ao}, {Bowler}, {Collier},
  {Deane}, {Delhaize}, {Frank}, {Glowacki}, {Hatfield}, {Maddox}, {Marchetti},
  {Matthews}, {Prandoni}, {Randriamampandry}, {Randriamanakoto}, {Smith},
  {Taylor}, {Thomas}, and {Vaccari}]{Whittam2022}
I.~H. {Whittam} et al.
\newblock \emph{\mnras}, 516\penalty0 (1):\penalty0 245--263, Oct. 2022.
\newblock \doi{10.1093/mnras/stac2140}.

\bibitem[{Zasowski} et~al.(2025){Zasowski}, {Jha}, {Chomiuk}, {Fan}, {Hickox},
  {Huber}, {Kerins}, {Kobulnicky}, {Lauer}, {Sako}, {Shapley}, {Stephens},
  {Weinberg}, and {Williams}]{Zasowski2025}
G.~{Zasowski} et al.
\newblock \emph{arXiv e-prints}, art. arXiv:2505.10574, May 2025.
\newblock \doi{10.48550/arXiv.2505.10574}.

\bibitem[{Zhang} et~al.(2024){Zhang}, {Liu}, and {Bromm}]{Zhang2024}
S.~{Zhang}, B.~{Liu}, and V.~{Bromm}.
\newblock \emph{\mnras}, 528\penalty0 (1):\penalty0 180--197, Feb. 2024.
\newblock \doi{10.1093/mnras/stad3986}.

\end{thebibliography}
